\pdfoutput=1
\documentclass{article}

\usepackage{iclr2027_conference}
\usepackage{times}
\usepackage{latexsym}
\usepackage{inconsolata}
\usepackage{enumitem}
\usepackage[utf8]{inputenc}
\usepackage[T1]{fontenc}
\usepackage{booktabs}
\usepackage{amsfonts}
\usepackage{amsmath}
\usepackage{nicefrac}
\usepackage{microtype}
\usepackage{xcolor}
\usepackage{graphicx}
\usepackage{wrapfig}
\usepackage{subcaption}
\usepackage{placeins}
\usepackage{multirow}
\usepackage{makecell}
\usepackage{listings}
\usepackage{hyperref}
\usepackage{url}

\newcommand{\para}[1]{\par\noindent\textbf{#1}\quad\ignorespaces}

\iclrfinalcopy

\title{Discovering Interpretable Acoustic Features \\in Autoregressive Voice-Cloning TTS Models}

\author{%
  Anuj Diwan \\
  University of Texas at Austin \\
  \texttt{anuj.diwan@utexas.edu} \\
  \And
  David Harwath \\
  University of Texas at Austin \\
  \texttt{harwath@utexas.edu} \\
  \And
  Eunsol Choi \\
  New York University \\
  \texttt{eunsol@nyu.edu} \\
}

\begin{document}

\maketitle

\begin{abstract}
Voice-cloning TTS models synthesize an unseen speaker's voice from brief speech prompts, but it remains unclear which acoustic properties they represent and use. We introduce a feature-based interpretability framework for autoregressive voice-cloning TTS models, analyzing features from sparse autoencoders (SAEs), PCA, and random directions via sparse probing, steering, and quantitative auto-interpretability. We focus on Qwen3-TTS and extend probing and steering to $5$ additional models to test generalization across architectures.
These models encode concepts like pitch, energy, noise, and accent, but sparse prediction and steering are effective only for some concepts. Strong probing does not always translate to steering or vice-versa. SAEs are useful for sparse probing of human-labelled concepts and novel-concept discovery, while PCA is competitive for dense probing and steering. Auto-interpretability finds $684$ automatically validated SAE features in Qwen3-TTS covering speaker clusters, speaking style, and recording conditions beyond known concepts. Overall, we perform a comprehensive interpretability study of autoregressive voice-cloning TTS models.
\end{abstract}

\section{Introduction}
\label{sec:intro}

Zero-shot voice-cloning TTS models~\citep{chen2025neuralcodeclanguagemodels,du2024cosyvoice2scalablestreaming,Qwen3-TTS} synthesize speech in an unseen speaker's voice from a few seconds of prompt speech. Speech entangles speaker identity, vocal texture, emotion, and pacing. We ask which concepts these models represent, whether they can be predicted from a few features and steered during generation, and what their features capture beyond predefined labels. We study these questions through sparse probing, generation-time steering, and quantitative auto-interpretability for novel concept discovery.
This provides a way to test acoustic controls in TTS models and inspect speech properties beyond predefined labels.

Our analysis builds on sparse-autoencoder (SAE) work~\citep{cunningham2023sparseautoencodershighlyinterpretable,gao2024scalingevaluatingsparseautoencoders} that decomposes activations into interpretable feature dictionaries. Earlier speech SAE work~\citep{aparin2026audiosaeunderstandingaudioprocessingmodels} studies encoders, while earlier TTS work~\citep{varshavskyhassid2024semanticlatentspacediffusionbased,lenglet2025closerlook} studies known concepts in non-voice-cloning models. We instead study known and novel concepts in zero-shot voice-cloning models. The closest speech auto-interpretability work, AudioSAE~\citep{aparin2026audiosaeunderstandingaudioprocessingmodels}, uses natural-language labels for qualitative inspection; we quantitatively test whether labels are coherent, predictive of held-out audio, and stable under relabeling. While CoCoEmo~\citep{wang2026cocoemo} uses supervised steering directions and controllable TTS systems~\citep{diwan2025scalingrichstylepromptedtexttospeech,du2024cosyvoice2scalablestreaming} use explicit style labels, we steer using unsupervised feature dictionaries. While contemporaneous work on CosyVoice3~\citep{koriagin2026interpretingsteeringtexttospeechlanguage} also performs SAE probing and steering, we compare feature methods across $6$ TTS models to test generalization across architectures. Rather than $3$ select concepts, our steering evaluation covers $55$ extractor concepts.

\begin{figure*}[ht]
    \centering
    \includegraphics[width=\textwidth]{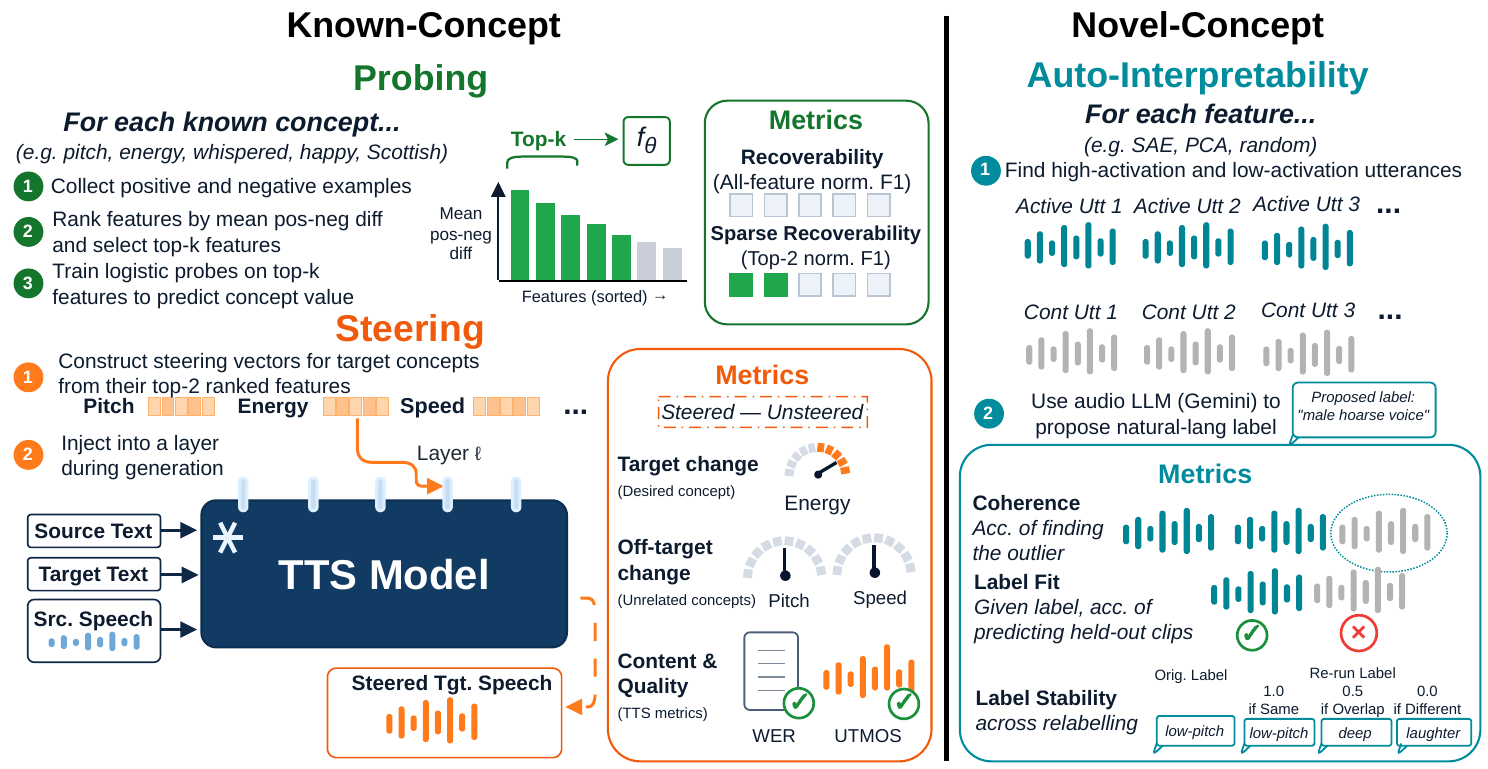}
    \caption{Our main interpretability evaluation pipeline. We evaluate feature sets via known-concept probing and steering across $6$ voice-cloning TTS models, and through quantitative auto-interpretability in Qwen3-TTS for novel concept discovery.}
    \label{fig:mainfig}
    \vspace{-0.5em}
\end{figure*}

We study Qwen3-TTS~\citep{Qwen3-TTS} as our primary model, and extend probing and steering to $5$ additional Fish, Higgs, and CosyVoice2 models to test generalization across architectures. We convert activations into utterance-level features using SAE, PCA, random directions, and untrained SAEs. We find that models encode pitch, energy, noise, accent, and emotion, but only some concepts are predictable from a few features. PCA and random directions are strong baselines, while SAEs outperform PCA on average in sparse prediction of human-labelled concepts. Feature directions can steer pitch, energy, noise, and spectral properties, with effects varying by model and concept. Strong sparse prediction does not always translate to steering or vice-versa. Auto-interpretability yields $684$ automatically validated SAE features in Qwen3-TTS spanning shared voice profiles, speaking styles, and recording conditions beyond known concepts.

Future studies could use our probing setup to track representation changes during fine-tuning. Discovered features could also guide investigations of unwanted transfer of background music or reverberation from reference speech.
Steering demos and discovered features are at \url{https://anonymous.4open.science/w/tts_interp_demo/}. \ificlrfinal Code will be released at \url{https://github.com/ajd12342/voice-cloning-interp}.\fi

In summary, our contributions are:
\begin{itemize}[leftmargin=*, itemsep=1pt, topsep=2pt]
\item We introduce a feature-based interpretability framework for zero-shot voice-cloning TTS, comparing SAE, PCA, and random features through sparse probing and generation-time steering. We probe $55$ extractor-derived and $21$ human-labelled concepts and steer all extractor concepts.
\item We test the generalizability of our pipeline and findings across architectures, extending probing and steering from our detailed Qwen3-TTS study to $5$ additional Fish, Higgs, and CosyVoice2 models.
\item We develop a quantitative auto-interpretability pipeline that tests labeled features for coherence, predictive fit, and stability, discovering $684$ automatically validated SAE features in Qwen3-TTS spanning voice profiles, speaking styles, and recording conditions.
\end{itemize}

\section{Models, Features, and Data}
\label{sec:features}

We use Qwen3-TTS-12Hz-0.6B-Base~\citep{Qwen3-TTS} (Qwen3-TTS) for our main probing, steering, and novel-concept analyses. We extend probing and steering to Fish S1-mini~\citep{liao2024fishspeechleveraginglargelanguage}, Fish S2 Pro~\citep{liao2026fishaudios2technical}, Higgs TTS 2~\citep{higgsaudio2025}, Higgs TTS 3~\citep{bosonai_higgs_audio_tts_v3_2026}, and CosyVoice2-0.5B~\citep{du2024cosyvoice2scalablestreaming} to test the generalization of our pipeline.

\begin{figure}
    \centering
    \includegraphics[width=\linewidth]{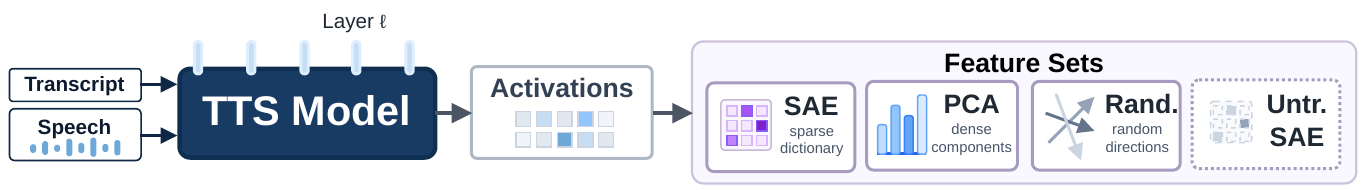}
    \caption{Feature extraction from TTS residual-stream activations using SAE, PCA, random directions, and untrained SAEs. We study multiple layers in Qwen3-TTS and one in other models.}
    \label{fig:feature_extraction}
    \vspace{-1em}
\end{figure}

Figure~\ref{fig:feature_extraction} shows how we convert residual-stream activations into utterance-level features using TopK SAEs~\citep{gao2024scalingevaluatingsparseautoencoders}, PCA, random directions, and untrained SAEs.

\subsection{Models and Activation Extraction}
\label{sec:model_arch}
We restrict analysis to the autoregressive backbone residual stream in each model. During inference, we prefill each model with the reference text, target text and reference speech, autoregressively generate target speech, and finally use their respective decoders to decode the generated tokens into a waveform akin to other autoregressive TTS models~\citep{du2024cosyvoice2scalablestreaming,chen2025neuralcodeclanguagemodels}.

\para{Qwen3-TTS}
We select Qwen3-TTS for its architectural compatibility with text SAEs~\citep{gao2024scalingevaluatingsparseautoencoders}, competitive TTS performance~\citep{Qwen3-TTS}, and lack of explicit style supervision. We use the $0.6$B variant, which represents speech with $16$ codebooks at $12$Hz. Its $28$-layer backbone with hidden size $d=1024$ predicts the first-codebook token for each speech frame, and a $5$-layer multi-token prediction (MTP) module predicts the remaining $15$ codebook tokens.

\para{Other Models}
\textbf{Fish S1-mini} is a $0.5$B model with a $28$-layer backbone and hidden size $d=1024$, while \textbf{Fish S2 Pro} has a $4$B backbone with $36$ layers and $d=2560$, followed by a $400$M fast transformer. Both use $10$ codec streams, the autoregressive backbone predicts the first stream, and a fast transformer predicts the remaining $9$. \textbf{Higgs TTS 2} has $28$ backbone layers with $d=3072$ and $5.8$B parameters: $3.6$B in the language model and $2.2$B in audio-specific feed-forward layers. \textbf{Higgs TTS 3}~\citep{bosonai_higgs_audio_tts_v3_2026} uses a $4$B backbone with $36$ layers and $d=2560$. Both predict all $8$ codec streams directly in the backbone, with the streams offset across generation steps. \textbf{CosyVoice2} uses a $0.5$B autoregressive language model with $24$ layers and $d=896$ to predict a single stream of semantic speech tokens. A conditional flow model converts these into mel features, followed by a vocoder. The flow model also receives reference speech tokens and reference mel features.

\para{Activation Extraction}
For Qwen3-TTS, we tokenize each utterance's transcript and observed speech, run a teacher-forced forward pass, and extract speech-region activations from layers $\{3, 9, 14, 19, 26\}$ (early, $\tfrac{1}{3}$-depth, $\tfrac{1}{2}$-depth, $\tfrac{2}{3}$-depth, near-final). For the additional models, we extract activations at only one layer, approximately the same relative depth as Qwen3-TTS layer $19$: layer $19$ for Fish S1-mini and Higgs TTS 2, $24$ for Fish S2 Pro and Higgs TTS 3, and $16$ for CosyVoice2. We use stride-$3$ token subsampling for feature fitting and no subsampling otherwise.

\subsection{Corpora and Splits}
\label{sec:corpora_splits}

\para{Main Corpus}
We use a $3.3$k-hr \textit{main corpus} ($1.5$M utterances, $151.2$k speaker IDs) spanning clean audiobooks, in-the-wild speech, accents and emotional data drawn from LibriTTS-R~\citep{koizumi2023librittsrrestoredmultispeakertexttospeech}, MSP-Podcast~\citep{lotfianmsppodcast2017}, Emilia~\citep{he2024emiliaextensivemultilingualdiverse}, VoxCeleb~\citep{nagrani2020voxceleb} and more. We split it $85$\%/$15$\% by speaker ID into disjoint \textit{feature-fitting} and \textit{analysis} splits containing $1.2$M and $218.4$k utterances, respectively. The feature-fitting split trains SAEs, fits PCA, and instantiates the remaining feature sets; the analysis split is used during evaluation. We use the full feature-fitting split for Qwen3-TTS and CosyVoice2, and a $20\%$ random subset for the Fish and Higgs models due to compute constraints.

\para{Other Evaluation Corpora}
The \textit{ParaSpeechCaps tag concept set} is built from ParaSpeechCaps-Base~\citep{diwan2025scalingrichstylepromptedtexttospeech} and provides human-labelled annotations for $102.4$k utterances from $566$ speakers, filtered to $21$ concepts that have enough speakers. The \textit{steering eval set} contains $100$ source utterances from $32$ speakers and $100$ target transcripts sampled from LibriTTS-R~\citep{koizumi2023librittsrrestoredmultispeakertexttospeech} test-clean. Full corpus statistics and details are in Appendix~\ref{app:data_details}.

\subsection{Feature Sets}
\label{sec:feature_sets}

\para{SAE}
We train a TopK SAE~\citep{gao2024scalingevaluatingsparseautoencoders} for each studied layer on activations from the feature-fitting split. Each SAE learns an $M$-feature dictionary with at most $K$ active features per token. For Qwen3-TTS, we sweep $M$ and $K$ to balance reconstruction quality and the fraction of inactive features, selecting $M=\{d,2d,2d,4d,12d\}$ for layers $\{3,9,14,19,26\}$ and $K=96$ throughout. For Fish, Higgs and CosyVoice2, we fix $M=4d$ and $K=96$ following Qwen3-TTS layer $19$ without separate sweeps. We train with Adam, learning rate $5\times10^{-4}$, and batches of $4096$ token activations. Full training settings and sweep results are in Appendices~\ref{app:sae_details} and~\ref{app:sae_sweeps}. We max-pool each feature over tokens to obtain utterance-level features, as in~\citet{aparin2026audiosaeunderstandingaudioprocessingmodels}.

\para{Other Feature Sets}
For \textbf{PCA}, we fit principal components on mean-pooled activations from a $20\%$ random subset of the feature-fitting split. For \textbf{random directions}, we project mean-pooled activations onto random unit vectors, matching the corresponding SAE width at each layer. We also use features from \textbf{untrained SAEs} with the same architecture as the trained SAEs.

\section{Probing and Steering Known Concepts}
\label{sec:known_concept}

Figure~\ref{fig:mainfig} summarizes our evaluation of predefined acoustic and paralinguistic concepts, inspired by~\citet{wu2025axbench}. We compare prediction from a few features with the full feature set using~\citet{karvonen2025saebench}'s sparse top-$k$ probing, then steer aligned features to test whether they causally affect generated speech without changing other concepts.

\subsection{Known-Concept Labels}
\label{sec:concept_extraction}

\para{Extractor Concepts}
We run extractors for pitch, energy, voicing, timing, noise, formants, emotion dimensions, and gender over the main corpus analysis split. We apply reliability filters, such as minimum voiced frames, so each concept is evaluated only when well-defined, and average frame-level outputs when needed. These extractors are also run on generated waveforms during steering.

\para{Human-Labelled Concepts}
We use the \textit{ParaSpeechCaps tag concept set} defined in Section~\ref{sec:corpora_splits}. Its annotations cover speaker and style concepts including vocal texture, clarity, loudness, emotion, and accent. These tags are used for sparse probing but not for steering because reliably assigning them to generated waveforms would require collecting human annotations.

\subsection{Sparse Probing Setup}
\label{sec:sparse_topk_method}

\para{Binary Dataset Setup}
We convert extractor and human labels into binary concepts. Continuous-valued concepts such as pitch become low- and high-tail concepts; probabilistic concepts, such as emotion, stay only high-tail; and human-labelled tags become binary concepts with tagged utterances as positives. We split each dataset $70$\%/$10$\%/$20$\% by speaker ID into disjoint train, val, and test splits for probing. This yields $55$ extractor concepts and $21$ final human-labelled concepts. More details in Appendix~\ref{app:known_concept_extractors}.

\para{Feature Selection and Probe Training}
For every concept, feature set, and layer, we rank features by the absolute positive-negative mean difference on the train split. We then train an $\ell_2$-regularized logistic probe on the top $k \in \{1,2,5,10,\mathrm{all}\}$ features, tune the decision threshold on val F1, and evaluate F1 on the test split. All models use the same splits for probe-training and held-out test sets. Comparisons across models use Qwen3-TTS layer $19$ and the layers specified in Section~\ref{sec:model_arch} for the other models. Additional Qwen3-TTS details are in Appendix~\ref{app:sparse_probing_details}.

\para{Evaluation Metrics}
We report normalized F1 to compare concepts with different positive rates. For a given concept, let $\mathrm{F1}_{\mathrm{null}}$ be the F1 of an always-positive classifier on the test split and define normalized $\mathrm{nF1}_k=(\mathrm{F1}_k-\mathrm{F1}_{\mathrm{null}})/(1-\mathrm{F1}_{\mathrm{null}})$.

\emph{Recoverability} is $\mathrm{nF1}_{k=\mathrm{all}}$, measuring how well the full feature set recovers the concept. \emph{Sparse recoverability} is $\mathrm{nF1}_{k=2}$, measuring how well the concept can be recovered from only $2$ features.

\subsection{Sparse Probing Results}
\label{sec:probing_results}
\begin{figure*}[!htb]
    \centering
    \begin{minipage}{0.9\linewidth}
    \begin{subfigure}[t]{0.5405\linewidth}
        \centering
        \vspace{0pt}
        \includegraphics[width=\linewidth]{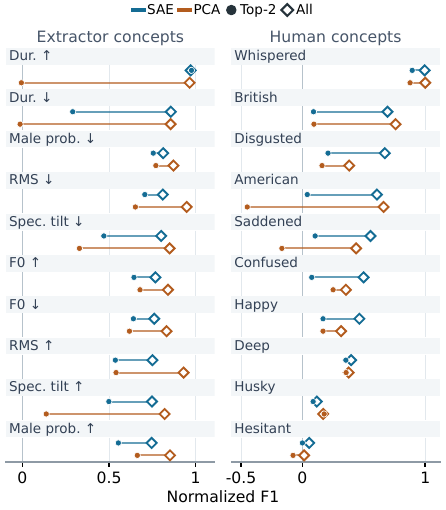}
        \caption{Qwen3-TTS layer $19$}
        \label{fig:concept_metrics}
    \end{subfigure}\hfill
    \begin{subfigure}[t]{0.4449\linewidth}
        \centering
        \vspace{0pt}
        \includegraphics[width=\linewidth]{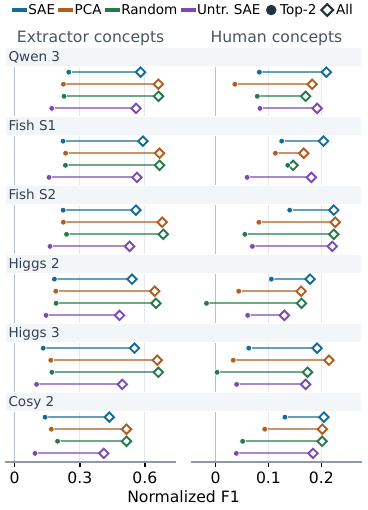}
        \caption{Six-model comparison}
        \label{fig:probing_across_models}
    \end{subfigure}
    \end{minipage}

    \caption{All-feature and top-$2$ nF1. (a) Top $10$ extractor and human-labelled concepts by SAE recoverability in Qwen3-TTS. \emph{Takeaway}: Many concepts, especially extractor concepts, are recoverable but few are sparsely predicted from $2$ features. (b) Mean nF1 over $55$ extractor and $21$ human-labelled concepts across $6$ models (layers in Section~\ref{sec:model_arch}). Full lists in Appendix~\ref{app:extractor_concept_metrics_appendix}. \emph{Takeaway}: PCA and random directions have higher all-feature extractor nF1 than SAE, while SAE has higher top-$2$ human-labelled nF1 than PCA across all six models.}
    \vspace{-0.5em}
    \label{fig:probing_summary}
\end{figure*}

Figure~\ref{fig:concept_metrics} reports all-feature and top-$2$ nF1 for the top $10$ extractor and human-labelled concepts by SAE recoverability at Qwen3-TTS layer $19$; full lists are in Appendix~\ref{app:extractor_concept_metrics_appendix}. Figure~\ref{fig:probing_across_models} compares all-feature and top-$2$ recoverability across all six models.

\textbf{Extractor concepts} are broadly recoverable, with duration, energy, pitch, spectral tilt, and gender among the strongest. Pitch, energy, and gender also have strong sparse recoverability, while speaking rate, accent, and many spectral properties are much better predicted using all features.

\textbf{Human-labelled concepts} are less consistently recoverable; only accents and some styles are. Whispered has strong SAE sparse recoverability across all models. British accent is more sparsely recoverable from SAEs in Fish and CosyVoice2 than at Qwen3-TTS layer $19$, although Qwen3-TTS improves at layer $26$ (Appendix~\ref{app:known_concept_layerwise}).

\textbf{Comparing feature sets}, we find that across models, PCA and random directions have higher extractor-concept recoverability than SAE, while SAE exceeds PCA in sparse recoverability of human-labelled concepts (Figure~\ref{fig:probing_across_models}). SAE leads in sparse extractor-concept recoverability only in Qwen3-TTS. Results comparing various $k$ values are in Appendix~\ref{app:known_concept_k}.

\textbf{Comparing models}, we find broadly similar all-feature nF1 for extractor concepts in Qwen3-TTS, Fish, and Higgs. With only $2$ features, Qwen3-TTS and Fish have higher mean nF1 than Higgs and CosyVoice2 (Figure~\ref{fig:probing_across_models}). CosyVoice2 also has lower all-feature nF1, particularly for energy and spectral concepts, although duration and speaking rate remain well predicted.

\textbf{Layerwise analysis} plots for Qwen3-TTS are in Appendix~\ref{app:known_concept_layerwise}. For SAE features, extractor concepts are already recoverable in early layers, but sparse recoverability is strongest later. Human-labelled concepts improve in both metrics from early to later layers.

\subsection{Steering Setup}
\label{sec:known_concept_steering_method}

Steering asks whether features associated with known concepts can causally affect generation. We steer all $55$ extractor concepts using their top-$2$ SAE, PCA, or random features. Like prior text~\citep{subramani-etal-2022-extracting,turner2024steering,rimsky-etal-2024-steering} and speech~\citep{wang2026cocoemo,xie2025emosteertts} steering, we intervene at inference rather than learning style conditioning through training~\citep{wang2018style,diwan2025scalingrichstylepromptedtexttospeech}.

\para{Generation Setup}
Each example pairs source audio and transcript with a target transcript. After prefilling with the source, we add $\lambda$ times a steering vector to one layer's residual stream during generation~\citep{turner2024steering}. Uniform steering compounded over time in Qwen3-TTS, so we linearly decay the intervention to zero over the expected generation length (Appendix~\ref{app:steering_details}). We generate unsteered and steered speech at $\lambda=1$ on the same $100$ examples across models, using Qwen3-TTS layers $19$ and $26$ and the other models' layers from Section~\ref{sec:model_arch}.

\para{Steering Vectors}
For SAE, PCA, and random directions, we sum the concept's signed top-$2$ unit directions, each scaled by its cached activation standard deviation. Prior work instead tunes steering strength~\citep{rimsky-etal-2024-steering,durmus2024steering,chalnev2024improving}. A supervised DiffMean topline~\citep{marks2024the} uses the positive-minus-negative mean activation, unit-normalized and scaled by the std along that direction (Appendix~\ref{app:steering_details}).

\para{Evaluation Metrics}
We compare extractor scores between steered and unsteered waveforms, normalizing paired changes by each concept's analysis-split $95$th--$5$th percentile range. Measurements and changes are not clipped, so scores can exceed $1$.

\emph{Target Change} measures intended change~\citep{rimsky-etal-2024-steering,durmus2024steering} as the median signed normalized change across utterances, positive in the requested direction.

\emph{Off-target change} measures side effects~\citep{durmus2024steering,chalnev2024improving} on \textit{protected concepts}: non-target concepts we aim to preserve, excluding those strongly correlated with the target on the analysis split. We take the median absolute normalized change across utterances for each protected concept, then the median across concepts (Appendix~\ref{app:steering_details}).

For \emph{Quality}, we report paired mean $\Delta$WER (percentage points) using Whisper large-v3~\citep{radford2023robust} and $\Delta$UTMOS using UTMOS~\citep{saeki22c_interspeech}.

\subsection{Steering Results}
\label{sec:steering_results}
\begin{figure*}[!htb]
    \centering
    \includegraphics[width=\linewidth]{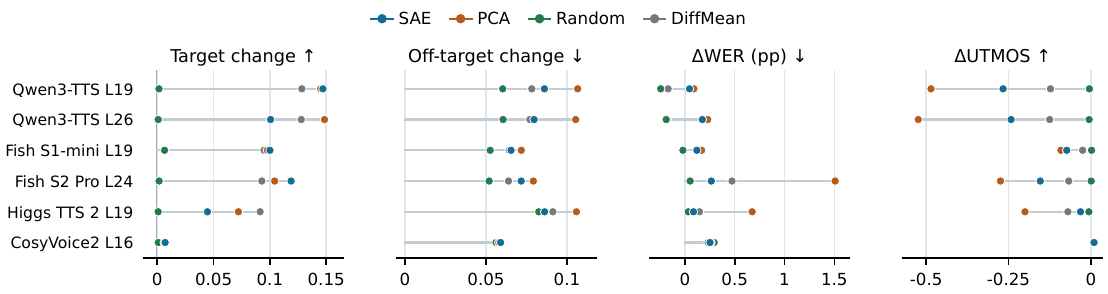}

    \caption{Steering at $\lambda=1$, averaged across $55$ extractor concepts on $100$ examples. Higgs TTS 3, where SAE steering severely degraded speech, is shown in Appendix~\ref{app:steering_results}. \emph{Takeaway}: In Qwen3-TTS and Fish, SAE target changes are competitive with PCA and supervised DiffMean, with smaller off-target changes than PCA; DiffMean has smaller UTMOS losses. Mean target changes are smaller in Higgs TTS 2 and CosyVoice2.}
    \vspace{-0.5em}
    \label{fig:steering_aggregate}
\end{figure*}

\begin{figure}[!htb]
    \centering
    \includegraphics[width=\linewidth]{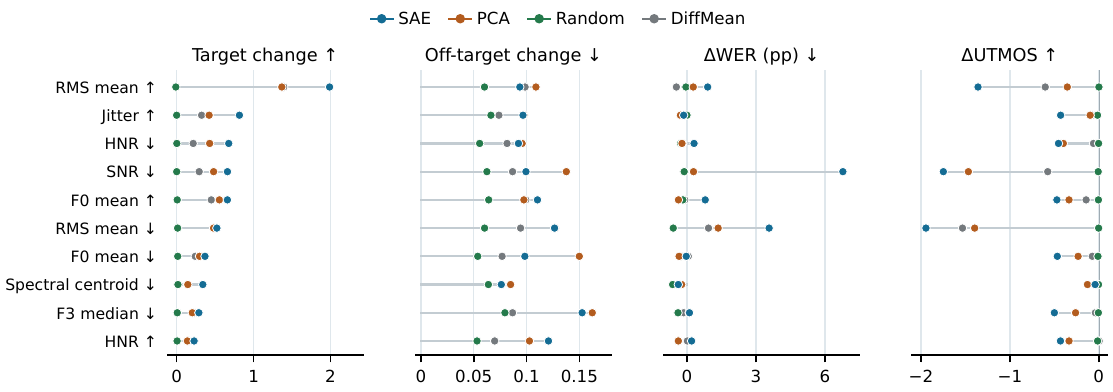}
    \caption{Top $10$ Qwen3-TTS layer-$19$ concepts by SAE median Target Change, comparing all four methods at $\lambda=1$ on $100$ examples. \emph{Takeaway}: SAE target changes are largest for energy, pitch, spectral and voice-quality properties. Quality losses partly reflect deliberately reducing SNR or pushing RMS toward either extreme.}
    \label{fig:steering_concept_lollipop}
    \vspace{-0.5em}
\end{figure}

Figure~\ref{fig:steering_aggregate} compares models, averaging metrics across all $55$ extractor concepts. Figure~\ref{fig:steering_concept_lollipop} shows the top $10$ concepts by SAE Target Change at Qwen3-TTS layer $19$ ($\lambda=1$), and Figure~\ref{fig:steering_strength_qwen} shows Qwen3-TTS strength sweeps. Expanded results are in Appendix~\ref{app:steering_results}.
We exclude human-labelled concepts because reliably evaluating them on generated speech would require new human annotations.

\textbf{Comparing feature sets}, we find that SAE, PCA, and DiffMean increase Target Change with strength. Stronger steering also increases off-target change and sometimes reduces quality (Figure~\ref{fig:steering_strength_qwen}). Random directions have near-zero Target Change despite competitive probing performance. In Qwen3-TTS and Fish, supervised DiffMean has the lowest off-target change, followed by SAE and PCA. SAE has smaller $\Delta$WER and $\Delta$UTMOS costs than PCA, while DiffMean has smaller UTMOS drops (Figure~\ref{fig:steering_aggregate}).

\textbf{Comparing concepts}, we find that steering works well for RMS, low-SNR, pitch, and spectral and voice-quality properties (Figure~\ref{fig:steering_concept_lollipop}). Low-SNR and RMS quality costs are partly expected: low-SNR steering should add noise, while RMS interventions deliberately make speech much louder or quieter. Spectral and voice-quality effects also appear in Fish. Most SAE changes in accent and emotion scores are small, while timing results are mixed.

\textbf{Comparing models}, we find larger mean Target Change from SAE, PCA, and DiffMean in Qwen3-TTS and Fish than in Higgs TTS 2 and CosyVoice2 (Figure~\ref{fig:steering_aggregate}). At the studied layers, SAE pitch steering is strong in Qwen3-TTS and Fish but weak in Higgs TTS 2, where PCA and DiffMean are more effective. SAE instead steers RMS and formants in Higgs TTS 2. In Higgs TTS 3, SAE severely degrades speech (Figure~\ref{fig:steering_multimodel_extended}).

\begin{figure*}[!htb]
    \centering
    \includegraphics[width=\linewidth]{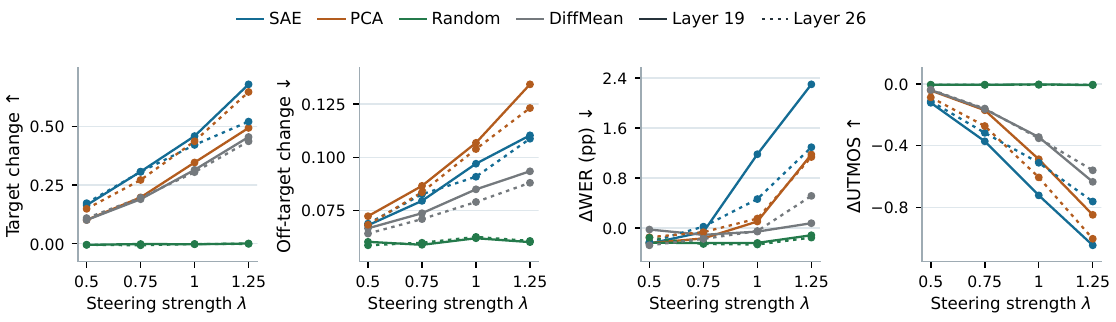}
    \caption{Qwen3-TTS strength sweeps, averaged across a fixed set of $9$ probing-selected targets on the same $100$ utterances (Appendix~\ref{app:steering_strength_sweep}). Layers $19$/$26$ use solid/dotted lines. \emph{Takeaway}: SAE, PCA, and DiffMean increase target change with strength while random directions remain ineffective. SAE has lower off-target change than PCA but pays a WER/UTMOS cost.}
    \label{fig:steering_strength_qwen}
\end{figure*}

Uniform steering in Qwen3-TTS can compound over time (e.g. progressively increasing pitch), motivating our decay schedule (Section~\ref{sec:known_concept_steering_method}). This anecdote suggests that the model may encode pitch locally rather than at the utterance-level, opening the possibility of temporally varying controls.

\textbf{Comparing probing and steering}, we find that strong probing does not always translate to steering. SAE sparse recoverability correlates weakly with Target Change across concepts in Qwen3-TTS, and barely at all in Fish (Appendix~\ref{app:probe_steering_association}). In Qwen3-TTS, top-$2$ SAE steering barely changes duration at layer $19$ despite strong sparse recoverability, but increases spectral centroid at layer $26$ despite weak sparse recoverability.

\section{Discovering Novel Concepts}
\label{sec:novel_concept}

Known-concept evaluations test whether model features can capture predefined concepts. To study model behavior beyond those, we auto-interpret~\citep{bills2023language} features leveraging an audio LLM to label features using highly activating audio examples, as depicted in Figure~\ref{fig:mainfig}. Due to API costs, we restrict this analysis to Qwen3-TTS. AudioSAE~\citep{aparin2026audiosaeunderstandingaudioprocessingmodels} uses natural-language labels qualitatively. We instead evaluate labeled features using three quantitative metrics: Coherence, Label Fit, and Label Stability.

\subsection{Auto-Interpretability Setup}

\para{Features and Utterances}
We randomly sample $500$ features per set at Qwen3-TTS layer $19$. For SAE features, we z-score nonzero activations, selecting utterances with $z>2$ as activating and activation $0$ as contrasts. For PCA and random directions, we z-score over all utterances and use $z>2$ as activating and $z\leq0$ as contrasts.

\para{Label Generation}
For each feature, we randomly sample $5$ activating and $5$ contrast utterances and ask Gemini $3.5$ Flash for the acoustic property that best separates them (Appendix~\ref{app:autointerp_details}). An example of a generated label is \textit{`Whispered or highly breathy, hushed speech'}.

\para{Label Evaluation}
Our metrics assess the coherence of activating utterances, the predictive fit of labels on held-out audio, and their stability across reruns and resampling. Since Gemini is used both to generate and evaluate labels, these automated metrics alone do not establish independent ground-truth validation, but jointly test Gemini's labeling abilities and feature-label alignment. They compare how consistently features can be described and predicted under a fixed audio LLM and protocol. We additionally measure human Label Fit on a subset of SAE features as an independent check (Section~\ref{sec:novel_results}).

\emph{Coherence} is the mean accuracy over $5$ trials where Gemini identifies the outlier among $10$ activating and $1$ contrast utterance, following intruder-style evaluation~\citep{paulo2026evaluating}.

\emph{Label Fit} is the accuracy when Gemini classifies held-out utterances ($10$ activating, $10$ contrast) as matching the label or not, following detection-based scoring~\citep{paulo2025automatically,bills2023language}.

\emph{Label Stability} compares labels regenerated from the original utterances and a disjoint set of $5$ activating and $5$ contrast utterances against the main label. Gemini scores each comparison as $1.0$ for same, $0.5$ for overlapping, and $0$ for different/unclear; we average the two scores.

\emph{Automatically validated features} have Coherence $\geq 3/5$, Label Fit $\geq 0.75$, and Label Stability $\geq 0.5$.

\para{Controls}
\emph{Random guessing} gives $\nicefrac{1}{11}$ Coherence, $0.5$ Label Fit, and $0.375$ Stability under uniform verdicts. \emph{Shuffled labels} tests Label Fit with another feature's label; \emph{random audio} tests Coherence and Label Fit with randomly sampled utterances.
\vspace{-5pt}
\subsection{Auto-Interpretability Results}
\label{sec:novel_results}

\begin{wraptable}{R}{0.5\linewidth}
    \caption{
    Auto-interpretability of $500$ Qwen3-TTS layer-$19$ features per set ($95\%$ bootstrap intervals). \emph{Takeaway}: SAE scores highest on all three metrics and yields the most validated features.
    }
    \label{tab:autointerp_results}
    \centering
    \footnotesize
    \setlength{\tabcolsep}{2.8pt}
    \renewcommand{\arraystretch}{0.95}
    \newcommand{\pmci}[2]{#1{\scriptscriptstyle\pm#2}}
    \newcommand{\saectrl}[1]{{\scriptsize$\hookrightarrow$}\,\emph{#1}}
    \resizebox{\linewidth}{!}{%
    \begin{tabular}{@{}lcccc@{}}
        \toprule
        \textbf{Feature Set} 
            & \textbf{Coherence}
            & \textbf{Label Fit}
            & \textbf{Label Stab.}
            & \textbf{Valid.} \\
        \midrule
        Rand. Guess 
            & $\pmci{0.09}{0.03}$ 
            & $\pmci{0.50}{0.04}$ 
            & $\pmci{0.38}{0.04}$ 
            & -- \\
        Rand. Dir. 
            & $\pmci{0.25}{0.02}$
            & $\pmci{0.60}{0.01}$
            & $\pmci{0.32}{0.02}$
            & $17$ \\
        PCA 
            & $\pmci{0.08}{0.01}$ 
            & $\pmci{0.50}{0.01}$
            & $\pmci{0.22}{0.02}$
            & $1$ \\
        SAE 
            & $\mathbf{0.43}{\scriptscriptstyle\pm0.03}$
            & $\mathbf{0.63}{\scriptscriptstyle\pm0.01}$
            & $\mathbf{0.41}{\scriptscriptstyle\pm0.03}$
            & $\mathbf{92}$ \\
        \saectrl{shuf. labels} 
            & -- 
            & $\pmci{0.50}{0.01}$
            & -- 
            & -- \\
        \saectrl{rand. audio} 
            & $\pmci{0.10}{0.01}$
            & $\pmci{0.50}{0.01}$
            & -- 
            & -- \\
        \bottomrule
    \end{tabular}
    }
\end{wraptable}

\textbf{Comparing feature sets}, we find that SAE scores highest on all three metrics and yields the most validated features ($92$, versus $17$ random and $1$ PCA; Table~\ref{tab:autointerp_results}). Random directions approach SAE on Label Fit but trail on Coherence and Stability, suggesting that Label Fit alone is insufficient for interpretability. PCA is near chance on Coherence and Label Fit. Shuffled-label and random-audio controls score near chance, indicating that the results depend on feature-matched labels and audio.

\para{Human evaluation}
We measure human Label Fit on $50$ validated and $50$ nonvalidated SAE features randomly sampled from the original $500$ layer-$19$ SAE features. For each feature, Amazon Mechanical Turk annotators judge whether its label matches each of $5$ activating and $5$ contrast utterances, with $2$--$3$ ratings per clip. Validated features achieve higher human Label Fit ($64.4\%$ vs.\ $51.0\%$, averaged across features), with nonvalidated features near the $50\%$ chance baseline. This suggests that validation selects labels that better distinguish activating from contrast utterances.

\para{Discovered concepts}
We manually grouped the $92$ validated SAE features by their primary label. Of these, $72$ are \emph{speaker cluster features} describing timbre and accent: $37$ male narration and synthetic-sounding reading voices, $14$ female voices spanning British narration, conversation and breathy storytelling, and $21$ other profiles including sports and podcast voices, gravelly male voices, and accented instruction. The remaining $20$ cover accent ($5$), laughter ($4$), recording conditions ($3$), whispering and synthetic-sounding speech ($2$ each), plus vocal fry, disfluencies, sports commentary, and expressive podcast delivery.

Since SAEs performed best, we extend discovery to the remaining eligible layer-$19$ features, yielding $592$ additional validated SAE features. These distinguish finer narration styles (deep audiobook voices, British academic reading, dramatic storytelling, robotic delivery), conversational profiles (raspy sports commentary, youthful gaming voices, gravelly lecturing), and laughter (during speech, amused delivery, background laughter). Other labels describe shouting, preaching, slow pacing, background music, reverberation, coughs, throat clearing, and backchannels.

\para{Speaker diversity analysis}
Auto-interpretability found many speaker cluster features, motivating further analysis. For each SAE feature, we extract speaker embeddings~\citep{speechbrain} and apply cosine agglomerative clustering (Appendix~\ref{app:autointerp_details}). Speaker clusters differ from speaker IDs: validated SAE features span a median of $26$ path-derived IDs across stored activating examples. Features with at most $5$ embedding clusters account for $19.5\%$ of eligible SAE features but $63.7\%$ of the $684$ validated features. This enrichment may reflect the substantial timbre capacity needed for voice-cloning, but also the auto-interpretability pipeline: timbre may be easier for the judge to perceive, describe, and verify, making such features more likely to pass all three checks.
\vspace{-5pt}

\section{Related Work}
\label{sec:related}
\para{SAEs for mechanistic interpretability}
SAEs learn overcomplete sparse dictionaries whose latents are often more interpretable than neurons~\citep{cunningham2023sparseautoencodershighlyinterpretable,bricken2023monosemanticity}. We follow recent work in evaluating their usefulness through sparse prediction, causal interventions, and steering~\citep{gao2024scalingevaluatingsparseautoencoders,karvonen2025saebench,wu2025axbench}.

\para{Interpreting speech and audio encoders}
SAEs, PCA, and related decompositions have been applied to speech/audio encoders~\citep{Chen_2022,hsu2021hubertselfsupervisedspeechrepresentation,defossez2022highfidelityneuralaudio}, revealing features aligned with pitch, emotion, accent, and loudness~\citep{aparin2026audiosaeunderstandingaudioprocessingmodels,mariotte2025sparseautoencodersmakeaudio,paek2025learninginterpretablefeaturesaudio,vanrensburg2026interpretingspeakercharacteristicsdimensions}. Some features can steer vocoders~\citep{paek2025learninginterpretablefeaturesaudio,vanrensburg2026interpretingspeakercharacteristicsdimensions} or reduce transcription errors~\citep{aparin2026audiosaeunderstandingaudioprocessingmodels}. Unlike this encoder-focused work, we study generative voice-cloning TTS models. Most of this work evaluates prespecified concepts. AudioSAE~\citep{aparin2026audiosaeunderstandingaudioprocessingmodels} uses natural-language labels for qualitative inspection; we instead evaluate labels quantitatively.

\para{TTS interpretability}
Prior TTS work finds PCA components for gender and intensity in Grad-TTS~\citep{varshavskyhassid2024semanticlatentspacediffusionbased} and probes encoder-decoder TTS models for syllabic and phonemic features~\citep{lenglet2025closerlook}. These studies focus on known concepts in non-voice-cloning models. We instead compare multiple feature sets through probing and steering of known concepts across $6$ zero-shot voice-cloning TTS models, and discover novel concepts in Qwen3-TTS. While contemporaneous work on CosyVoice3~\citep{koriagin2026interpretingsteeringtexttospeechlanguage} demonstrates steering with $3$ selected SAE features, we systematically compare SAE and PCA steering across all $55$ extractor targets, including off-target changes and speech quality. Our probing and steering comparisons span $6$ TTS models, testing generalization across architectures.

\para{Controllable TTS}
Most controllable TTS trains models on labels, style instructions, or reference audio~\citep{guo2023prompttts,yang2023instructtts,liu2023promptstyle,diwan2025scalingrichstylepromptedtexttospeech,du2024cosyvoice2scalablestreaming,chen2026flexivoice}. CoCoEmo and EmoSteer-TTS~\citep{wang2026cocoemo,xie2025emosteertts} use activation steering, but only along supervised emotion directions. Related emotion-sensitive neuron studies~\citep{zhao-etal-2026-discovering,zhao2026neuronlevelemotioncontrolspeechgenerative} also target predefined emotions in audio understanding and generation. Our steering experiments instead ask whether features from unsupervised dictionaries can serve directly as training-free controls for diverse acoustic concepts across frozen voice-cloning models.

\section{Conclusion}
\label{sec:conclusion}

We introduce a feature-based interpretability framework for voice-cloning TTS, combining probing and steering across $6$ models with novel-concept discovery in Qwen3-TTS. Many acoustic and paralinguistic concepts are recoverable, but only some are strongly predicted from a few features. Steering effects vary by model and concept, and strong sparse prediction does not always translate to effective steering. PCA and random directions are strong probing baselines, while SAEs are useful for sparse prediction of human-labelled concepts and novel-concept discovery. Auto-interpretability identifies $684$ automatically validated SAE features spanning speaker clusters, speaking style, and recording conditions in Qwen3-TTS. Future work should extend analysis to multilingual speech, study how features vary temporally, and explore composition of controls.

\ificlrfinal
\subsubsection*{Acknowledgements}
We thank Shauli Ravfogel and members of the UT NLP community for helpful feedback, and the Amazon Mechanical Turk annotators who participated in our human evaluation. We thank the Google Gemini Academic Program for providing Google Cloud credits.
\fi

\section*{AI Use Statement}
We used generative AI tools to implement methods and write code, refine experimental setups, and support qualitative analysis. All research ideas originated with the authors. Additionally, we used these tools to create plots, edit prose, fix LaTeX errors, and assist with literature searches. Gemini's role in feature labeling and evaluation is described in Section~\ref{sec:novel_concept}. We reviewed all AI-assisted work, checking results against saved outputs, citations against original papers, and inspecting plots, feature groupings, and prose. We take responsibility for the final text, claims, code and artifacts.

\bibliography{custom}
\bibliographystyle{iclr2027_conference}

\appendix

\section{Data Details}
\label{app:data_details}

Tables~\ref{tab:datasets}--\ref{tab:evaluation_corpora} report corpus and split statistics.
\begin{table}[!htbp]
    \centering
    \begin{minipage}[t]{0.48\linewidth}
    \vspace{0pt}
    \caption{Main corpus before splitting.}
    \label{tab:datasets}
\small
    \centering
    \begin{tabular}{lrrr}
        \toprule
        \textbf{Dataset} & \textbf{Utts} & \textbf{Hrs} & \textbf{IDs} \\
        \midrule
        LibriTTS-R & $230.7$k & $489.5$ & $2.3$k \\
        MSP-Podcast & $157.8$k & $247.9$ & $54.3$k \\
        Emilia & $766.0$k & $2.0$k & $88.8$k \\
        VCTK & $49.0$k & $55.4$ & $109$ \\
        VoxCeleb & $234.4$k & $500.0$ & $5.6$k \\
        ESD & $5.5$k & $5.4$ & $10$ \\
        RAVDESS & $1.4$k & $1.5$ & $24$ \\
        EARS & $11.4$k & $45.8$ & $75$ \\
        \midrule
        Total & $1.5$M & $3.3$k & $151.2$k \\
        \bottomrule
    \end{tabular}
    \end{minipage}\hfill
    \begin{minipage}[t]{0.48\linewidth}
    \vspace{0pt}
    \caption{Corpus split by speaker ID.}
    \label{tab:main_corpus_split}
\small
    \centering
    \begin{tabular}{lrrr}
        \toprule
        \textbf{Split} & \textbf{Utts} & \textbf{Hrs} & \textbf{IDs} \\
        \midrule
        Feature-fitting & $1.2$M & $2.8$k & $128.1$k \\
        Analysis & $218.4$k & $506.1$ & $23.2$k \\
        \midrule
        Total & $1.5$M & $3.3$k & $151.2$k \\
        \bottomrule
    \end{tabular}
    \par\medskip
    \caption{Held-out evaluation corpora.}
    \label{tab:evaluation_corpora}
\small
    \centering
    \begin{tabular}{lrrr}
        \toprule
        \textbf{Dataset} & \textbf{Utts} & \textbf{Hrs} & \textbf{IDs} \\
        \midrule
        PSC tags & $102.4$k & $224.3$ & $566$ \\
        Steering & $100$ & $0.22$ & $32$ \\
        \bottomrule
    \end{tabular}
    \end{minipage}
    \par\smallskip
    {\footnotesize\raggedright Sources: LibriTTS-R~\citep{koizumi2023librittsrrestoredmultispeakertexttospeech}, MSP-Podcast~\citep{lotfianmsppodcast2017}, Emilia~\citep{he2024emiliaextensivemultilingualdiverse}, VCTK~\citep{yamagishi2019cstrvctkcorpus}, VoxCeleb~\citep{nagrani2020voxceleb}, ESD~\citep{zhou2022emotionalvoiceconversiontheory}, RAVDESS~\citep{livingstone2018ryerson}, and EARS~\citep{richter2024earsanechoicfullbandspeech}.\par}
\end{table}

\para{Speaker IDs.}
Tables~\ref{tab:datasets}--\ref{tab:evaluation_corpora} count corpus-specific IDs. LibriTTS-R, VCTK, VoxCeleb, EARS, ESD, and RAVDESS use their speaker or actor IDs. Emilia uses batch-local speaker IDs. MSP-Podcast uses annotated speaker IDs when available and a separate utterance ID otherwise. Splits keep each ID together.

\para{ParaSpeechCaps tag concept set.}
We build the ParaSpeechCaps tag concept set from ParaSpeechCaps-Base~\citep{diwan2025scalingrichstylepromptedtexttospeech}. We first filter tags that have fewer than $500$ positive examples, leaving $35$ concepts. After splitting $70\%$/$10\%$/$20\%$ by speaker ID into disjoint train/val/test for probing, we filter any concepts that have at most $5$ positive test speakers. This leaves a final list of $21$ concepts, listed in Appendix~\ref{app:known_concept_extractors}.

\FloatBarrier
\section{Model and SAE Training Details}
\label{app:sae_details}

\para{TopK SAE.}
For each studied layer, we train a TopK SAE~\citep{gao2024scalingevaluatingsparseautoencoders} on speech-region residual-stream token activations. Given an activation $\mathbf{x} \in \mathbb{R}^d$, the SAE computes
\begin{align}
    \mathbf{z} &= W_{\mathrm{enc}}(\mathbf{x} - \mathbf{b}_{\mathrm{dec}}) + \mathbf{b}_{\mathrm{enc}}, \\
    \mathbf{f} &= \mathrm{ReLU}\bigl(\mathrm{TopK}(\mathbf{z})\bigr), \\
    \hat{\mathbf{x}} &= W_{\mathrm{dec}}^\top \mathbf{f} + \mathbf{b}_{\mathrm{dec}},
\end{align}
where $\mathrm{TopK}(\cdot)$ keeps the $K$ largest values in $\mathbf{z}$ and zeros the rest, $W_{\mathrm{enc}},W_{\mathrm{dec}} \in \mathbb{R}^{M \times d}$ are the encoder and decoder weights, each row $W_{\mathrm{dec},j}$ is a decoder direction, and $M$ is the dictionary size. Hard sparsity via $\mathrm{TopK}$ replaces the $L_1$ penalty used in earlier SAE variants~\citep{cunningham2023sparseautoencodershighlyinterpretable}.

\para{Objective and dead-feature loss.}
We minimize reconstruction error with an auxiliary dead-feature loss following Gao et al.~\citep{gao2024scalingevaluatingsparseautoencoders}:
\begin{equation}
    \mathcal{L} = \|\mathbf{x} - \hat{\mathbf{x}}\|_2^2 + \alpha \mathcal{L}_{\mathrm{aux}},
\end{equation}
with $\alpha = \nicefrac{1}{32}$. A feature is considered dead if it has not activated in the last $5000$ training steps. The auxiliary loss asks dead features to reconstruct the current residual $\mathbf{x}-\hat{\mathbf{x}}$, giving them gradient signal on variance not explained by live features.

\para{Qwen3-TTS training setup.}
We hold out $1$\% of utterances for validation and train each SAE for $30$ epochs. A preliminary streaming pass samples up to $400$k train tokens for preprocessing statistics. We use Adam with batch size $4096$, learning rate $5 \times 10^{-4}$, and a linear warmup of $5000$ steps. We otherwise follow common SAE training practice~\citep{bricken2023monosemanticity,karvonen2025saebench}: decoder directions are projected to unit norm after every optimizer step, the component of $\nabla W_{\mathrm{dec}}$ parallel to each decoder direction is removed before the optimizer step, and encoder weights are initialized to match decoder directions. We normalize inputs by a per-layer scalar $s$ chosen so that $\mathbb{E}[\|s\mathbf{x}\|_2^2] = d$. The decoder bias $\mathbf{b}_{\mathrm{dec}}$ is initialized to the geometric median of the same train-token statistics pool.

\para{Other models' training setup.}
We use $M=4d$, $K=96$, and train with Adam for $100000$ steps with batch size $4096$ token activations, learning rate $5\times10^{-4}$, and $2000$ warmup steps. We select the checkpoint with the best reconstruction-validation performance. Higgs TTS 3 uses the checkpoint at step $42000$ as its SAE training deteriorated later.

\para{Diagnostic metrics.}
We compute diagnostics on held-out activations $\{\mathbf{x}_i\}_{i=1}^N$ with reconstructions $\{\hat{\mathbf{x}}_i\}_{i=1}^N$ and SAE activations $\{\mathbf{f}_i\}_{i=1}^N$. Explained variance (EV) is
\begin{equation}
    \mathrm{EV}=1-\frac{\sum_i \|\mathbf{x}_i-\hat{\mathbf{x}}_i\|_2^2}{\sum_i \|\mathbf{x}_i-\bar{\mathbf{x}}\|_2^2},
\end{equation}
where $\bar{\mathbf{x}}$ is the held-out activation mean. For each feature $j$, its density is $p_j=\frac{1}{N} \sum_i \mathbf{1}[f_{ij}>0]$. Dead-feature fraction (DFF) is $\frac{1}{M}\sum_j \mathbf{1}[p_j=0]$.

\subsection{SAE Hyperparameter Sweeps}
\label{app:sae_sweeps}

\begin{figure*}[!htbp]
    \centering
    \includegraphics[width=0.49\linewidth]{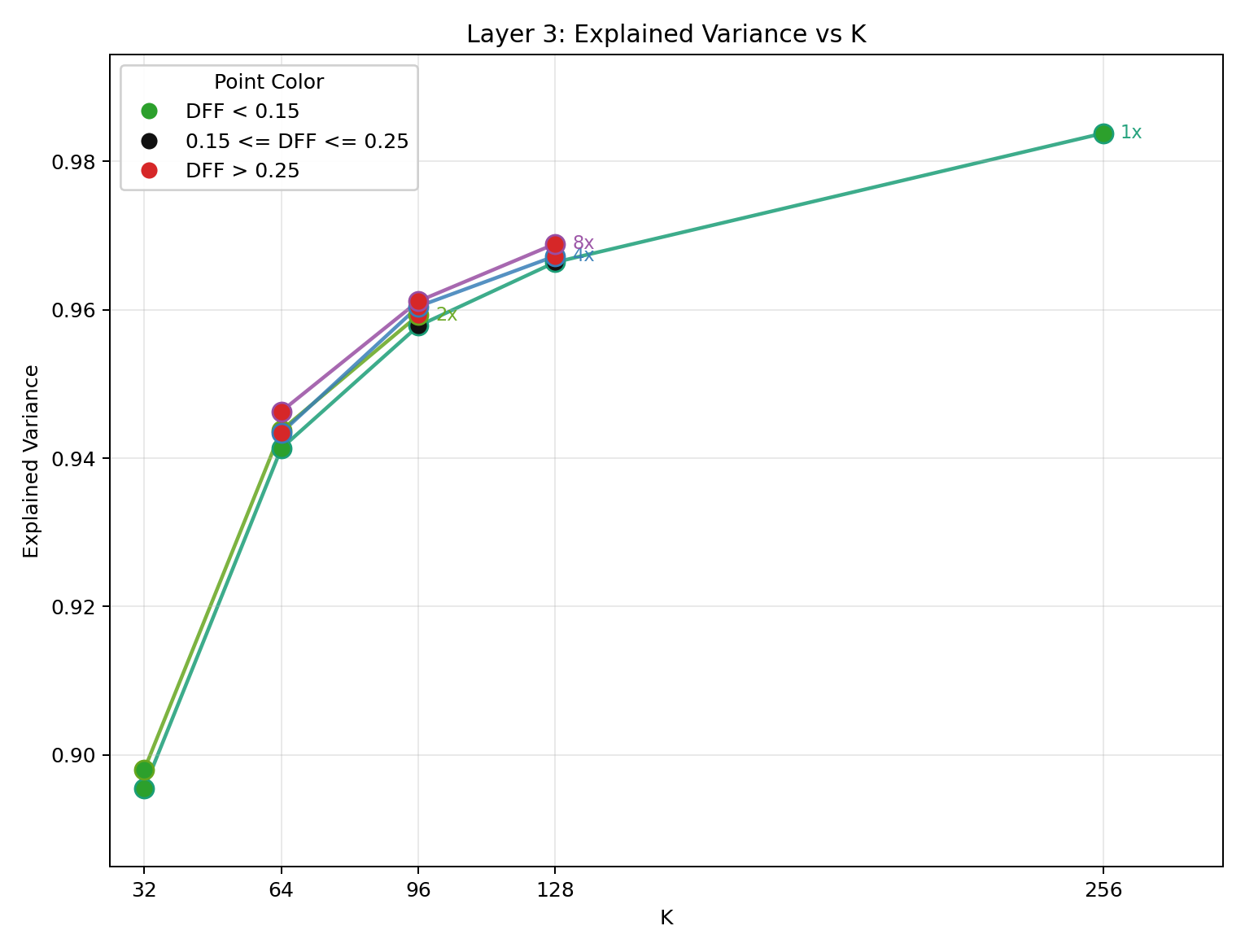}\hfill
    \includegraphics[width=0.49\linewidth]{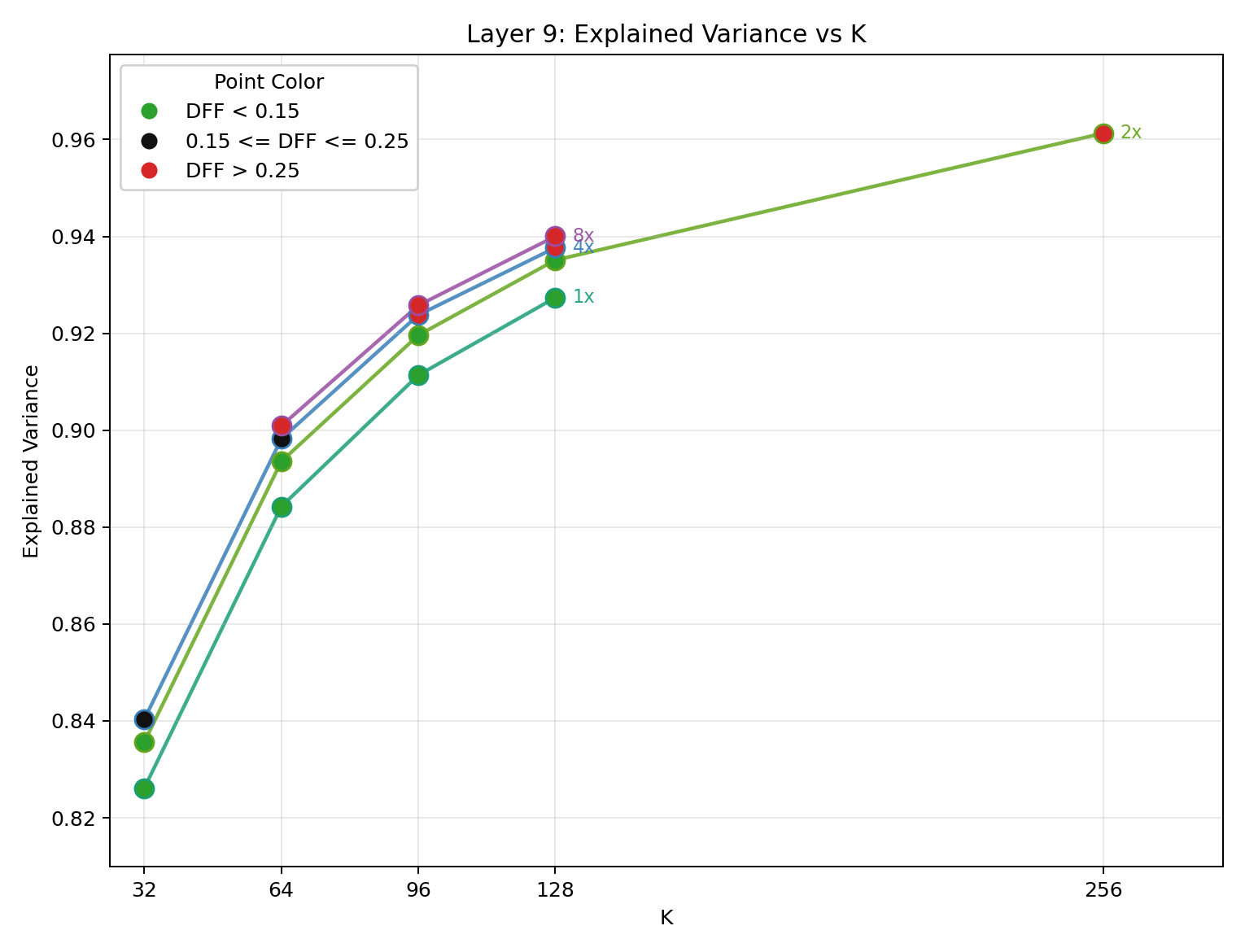}
    \par\smallskip
    \includegraphics[width=0.49\linewidth]{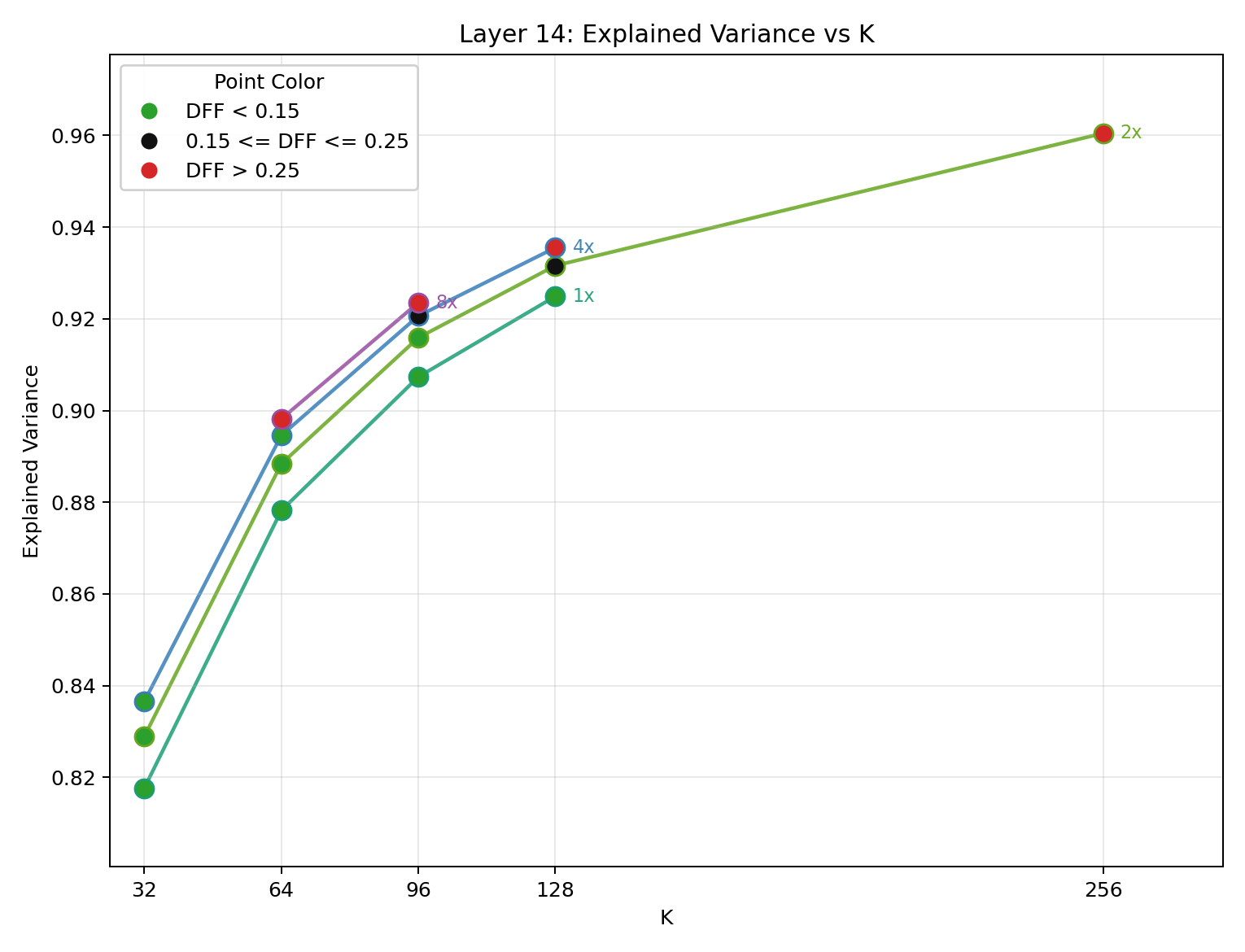}\hfill
    \includegraphics[width=0.49\linewidth]{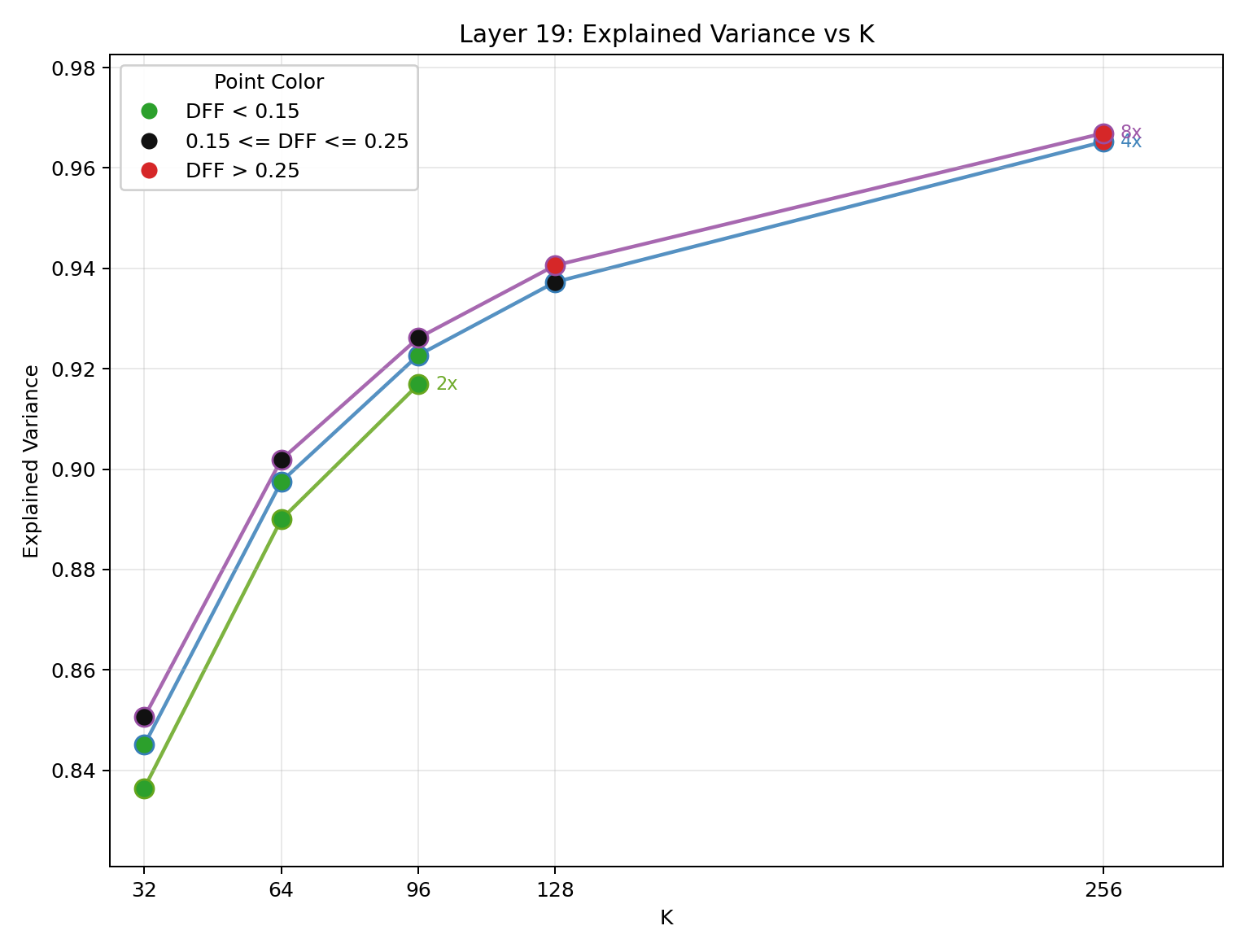}
    \par\smallskip
    \includegraphics[width=0.49\linewidth]{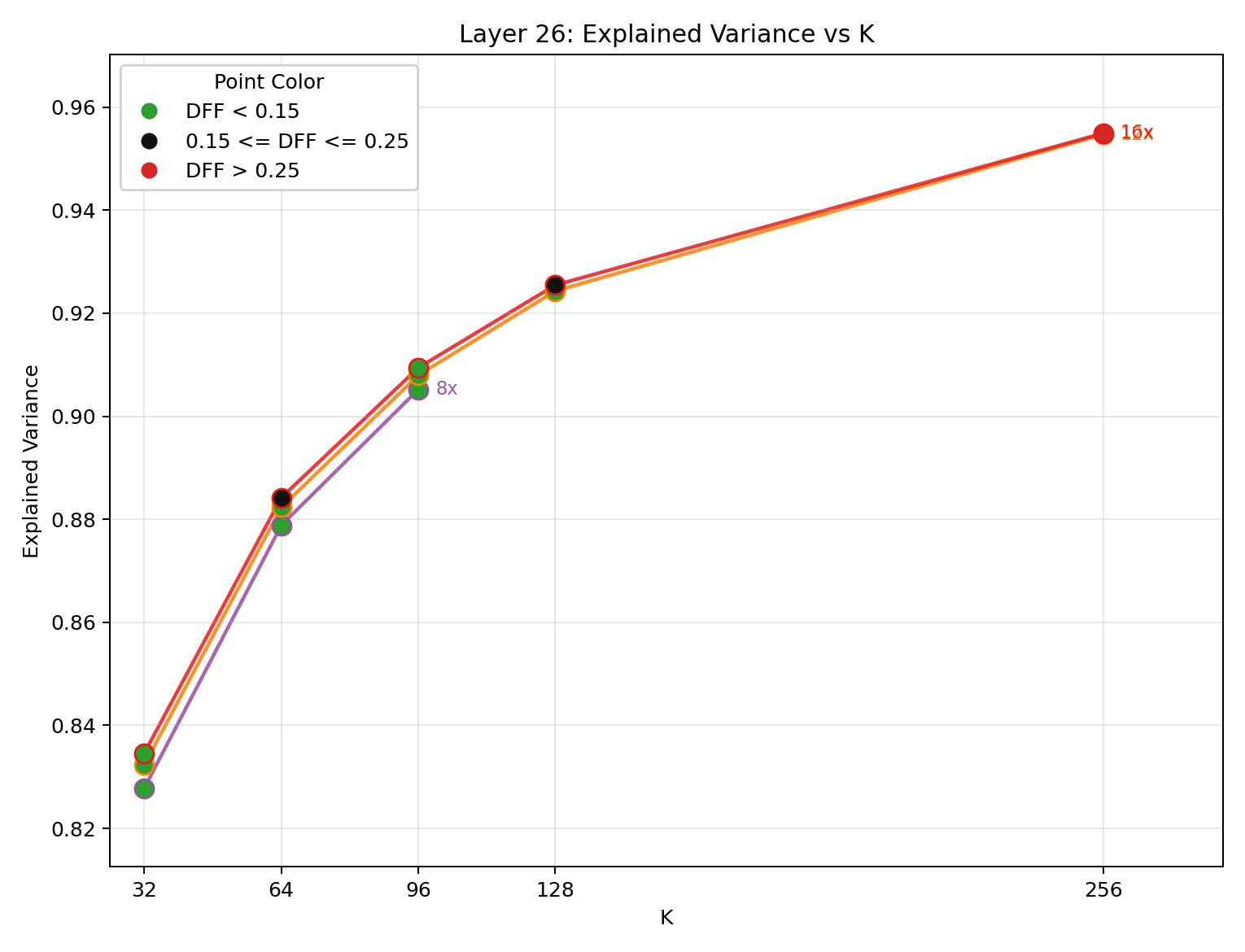}
    \caption{EV-vs-$K$ curves at epoch $10$ for layers $3$, $9$ (top), $14$, $19$ (middle), and $26$ (bottom). Lines vary $M$; point colors indicate DFF. We choose widths with DFF roughly below $0.15$ across $K\in\{32,64,96\}$, except at layer $3$, and select $K$ near the curve's shoulder.}
    \label{fig:sae_ev_dff_k_sweep}
\end{figure*}

We compare $10$-epoch Qwen3-TTS runs with widths $M \in \{d, 2d, 4d, 8d, 12d, 16d\}$ and $K \in \{32,64,96,128,256\}$; evaluated combinations are shown in Figure~\ref{fig:sae_ev_dff_k_sweep}. We choose the $M$ for which DFF remained roughly below $0.15$ across $K\in\{32,64,96\}$. Only layer $3$ did not cleanly satisfy this rule, so we simply selected the smallest $M=1024$ since larger $M$ made the DFF worse. This selected $M=\{d,2d,2d,4d,12d\}$ for layers $\{3,9,14,19,26\}$, showing that earlier layers prefer smaller dictionaries while later ones prefer wider dictionaries.

After choosing width, we selected $K$ from the EV-vs-$K$ curve. Figure~\ref{fig:sae_ev_dff_k_sweep} shows these curves for each probed layer. EV always rises with $K$, but DFF tends to jump once $K$ becomes too large. We use $K=96$ for all layers as it was a good shoulder point: it gives high EV while avoiding the larger DFF penalties seen at $K=128$ for some layers.

\FloatBarrier
\section{Known Concepts}
\label{app:known_concept_extractors}

Known-concept probing uses extractor concepts from audio and metadata extractors run on the main-corpus analysis split, and human-labelled concepts from the ParaSpeechCaps tag concept set described in Appendix~\ref{app:data_details}.

\para{Extractor inventory.}
Table~\ref{tab:known_concept_extractors} lists the automatic extractors and scalar values used before binarization. Each extractor writes scalar values and per-value reliability masks. Probing abstains on rows where an extractor marks a value unreliable, the value is missing, or the value is non-finite.

\begin{table*}[!htbp]
    \caption{Automatic concept extractors. Each scalar yields high- and low-tail concepts except emotion2vec and accent scores, which yield only high-tail concepts, giving $55$ targets in total.}
    \label{tab:known_concept_extractors}
    \small
    \centering
    \begin{tabular}{@{}p{0.16\textwidth}p{0.29\textwidth}p{0.48\textwidth}@{}}
    \toprule
    \textbf{Family} & \textbf{Extractor} & \textbf{Scalar outputs used} \\
    \midrule
    Pitch and voicing & PENN~\citep{penn2023} & \emph{f0 mean}, \emph{voiced fraction}, \emph{mean voiced run sec}, \emph{f0 slope} \\
    Energy and spectrum & librosa RMS and spectral centroid & \emph{rms mean}, \emph{rms slope}, \emph{spectral centroid mean} \\
    Timing & WebRTC VAD, g2p speaking rate, manifest duration & \emph{pause fraction}, \emph{mean pause sec}, \emph{speaking rate}, \emph{duration} \\
    Recording quality & Brouhaha~\citep{brouhaha_best_hf}; librosa STFT spectral tilt & \emph{snr}, \emph{c50}, \emph{spectral tilt} \\
    Voice quality & Praat/Parselmouth~\citep{parselmouth2018} & \emph{jitter local}, \emph{shimmer local}, \emph{hnr db} \\
    Formants & Praat/Parselmouth & \emph{f1 median}, \emph{f2 median}, \emph{f3 median} \\
    Gender & \href{https://huggingface.co/alefiury/wav2vec2-large-xlsr-53-gender-recognition-librispeech}{wav2vec2/XLS-R gender classifier} & \emph{gender male prob} \\
    Emotion & emotion2vec~\citep{emotion2vec2023}; \href{https://huggingface.co/audeering/wav2vec2-large-robust-12-ft-emotion-msp-dim}{MSP-DIM} & \emph{emo2vec happy}, \emph{emo2vec sad}, \emph{arousal}, \emph{valence} \\
    Accent & VoxProfile/Whisper~\citep{feng2025voxprofile} & \emph{accent east asia}, \emph{accent english}, \emph{accent germanic}, \emph{accent north america}, \emph{accent oceania}, \emph{accent romance}, \emph{accent scottish} \\
    \bottomrule
    \end{tabular}
\end{table*}

\para{Binarization.}
We convert scalar values into $55$ binary concepts. For scalars with both directions, we create high-tail and low-tail labels using the top and bottom $20$\% of reliable finite utterances. Positive examples are the corresponding tail, and negative examples are a random sample from reliable non-tail utterances, balanced to the number of positives. For emotion2vec and accent scores, we keep only high-confidence labels using the top $5$\% tail. Their negatives are high-confidence positives from other kept classes in the same family. We split each binary dataset $70$\%/$10$\%/$20$\% by speaker ID into disjoint train, val, and test splits before probing.

\para{Human-labelled concepts.}
The $21$ ParaSpeechCaps tags retained after filtering (Appendix~\ref{app:data_details}) are:
\begin{itemize}[leftmargin=*,nosep]
    \item Intrinsic: American, authoritative, booming, British, crisp, deep, flowing, hesitant, husky, loud, nasal, raspy, shrill, silky, slurred, and stammering.
    \item Situational: confused, disgusted, happy, saddened, and whispered.
\end{itemize}

\FloatBarrier
\subsection{Sparse Probing Details}
\label{app:sparse_probing_details}
Qwen3-TTS and CosyVoice2 additionally evaluate $k \in \{20,50,100,200\}$. Features that are nonzero on less than $0.1\%$ of probe-training utterances are excluded for probing. For each concept, feature set, and layer, features are ranked by the absolute positive-minus-negative mean activation difference. Ties are broken by feature index. After ranking, selected features are standardized using train means and stds before fitting.

Each probe is an $\ell_2$-regularized logistic regression fit with the sklearn liblinear solver, inverse regularization strength $C=1.0$, and at most $1000$ iterations. The decision threshold is selected on validation rows to maximize F1, with precision and then the smallest threshold possible used as tie-breakers.

\subsection{Steering Details}
\label{app:steering_details}

\para{Steering vectors.}
For SAE steering, the intervention vector for concept $c$ is
\[
\Delta_c(\lambda)=\lambda\sum_{j \in J_c} \sigma_j s_j W_{\mathrm{dec},j}/s_{\mathrm{in}},
\]
where $J_c$ contains the top $2$ sparse-probing features, $\sigma_j$ is the feature's cached nonzero activation standard deviation, $s_j \in \{-1,1\}$ is the train-split association sign, $W_{\mathrm{dec},j}$ is the SAE decoder direction, and $s_{\mathrm{in}}$ is the SAE input scale. PCA and random-direction steering use the same formula with the corresponding unit basis vector in place of $W_{\mathrm{dec},j}/s_{\mathrm{in}}$; these bases are unit-normalized before steering. The difference-of-means topline uses
\[
\Delta_c(\lambda)=\lambda\sigma_c v_c,
\]
where $v_c$ is the unit-normalized difference between positive and negative train-row mean hidden states and $\sigma_c$ is the train-row standard deviation along that axis.

\para{Qwen3-TTS strength sweep.}
\label{app:steering_strength_sweep}
We sweep $\lambda\in\{0.5,0.75,1.0,1.25\}$ at Qwen3-TTS layers $19$ and $26$ on $100$ examples. Both layers and all methods use the same $9$ targets selected across the original probing sweep: duration $\uparrow$, F0 mean $\uparrow/\downarrow$, male probability $\uparrow/\downarrow$, pause length $\uparrow$, RMS mean $\uparrow/\downarrow$, and SNR $\downarrow$.

\para{Temporal schedule.}
The intervention is applied only during generated-token steps, leaving the source-prompt prefill unmodified. For generated step $t$, we multiply the intervention by
\[
\alpha_t=\max(0, 1-t/T),
\]
where $T$ is the expected generated length estimated as the target/source transcript non-space character ratio times the number of source prompt codec tokens.

\para{Protected concepts.}
We exclude the target scalar and concepts with analysis-split absolute Spearman correlation $\geq 0.35$ with the target. High/low labels of the same scalar count as one protected property.

\para{Metric aggregation.}
For model comparisons and strength sweeps, we average available per-concept scores with equal weight, treating high/low directions separately. Missing scores are excluded from each metric's average. Each protected property uses pairs with valid target and property measurements. Off-target change is omitted if any protected property has no valid pairs. Within each concept, $\Delta$WER and $\Delta$UTMOS are paired means across utterances, with WER in percentage points.

\para{Probing and steering.}
\label{app:probe_steering_association}
Across $55$ targets, SAE top-$2$ nF1 and Target Change have Spearman $\rho=0.20/0.22$ at Qwen3-TTS layers $19/26$, and $-0.07/0.00$ in Fish S1/S2.

\FloatBarrier
\newpage
\noindent\begin{minipage}{\linewidth}
\subsection{Additional Steering Results}
\label{app:steering_results}
    \centering
    \captionsetup{hypcap=false}
    \includegraphics[width=\linewidth]{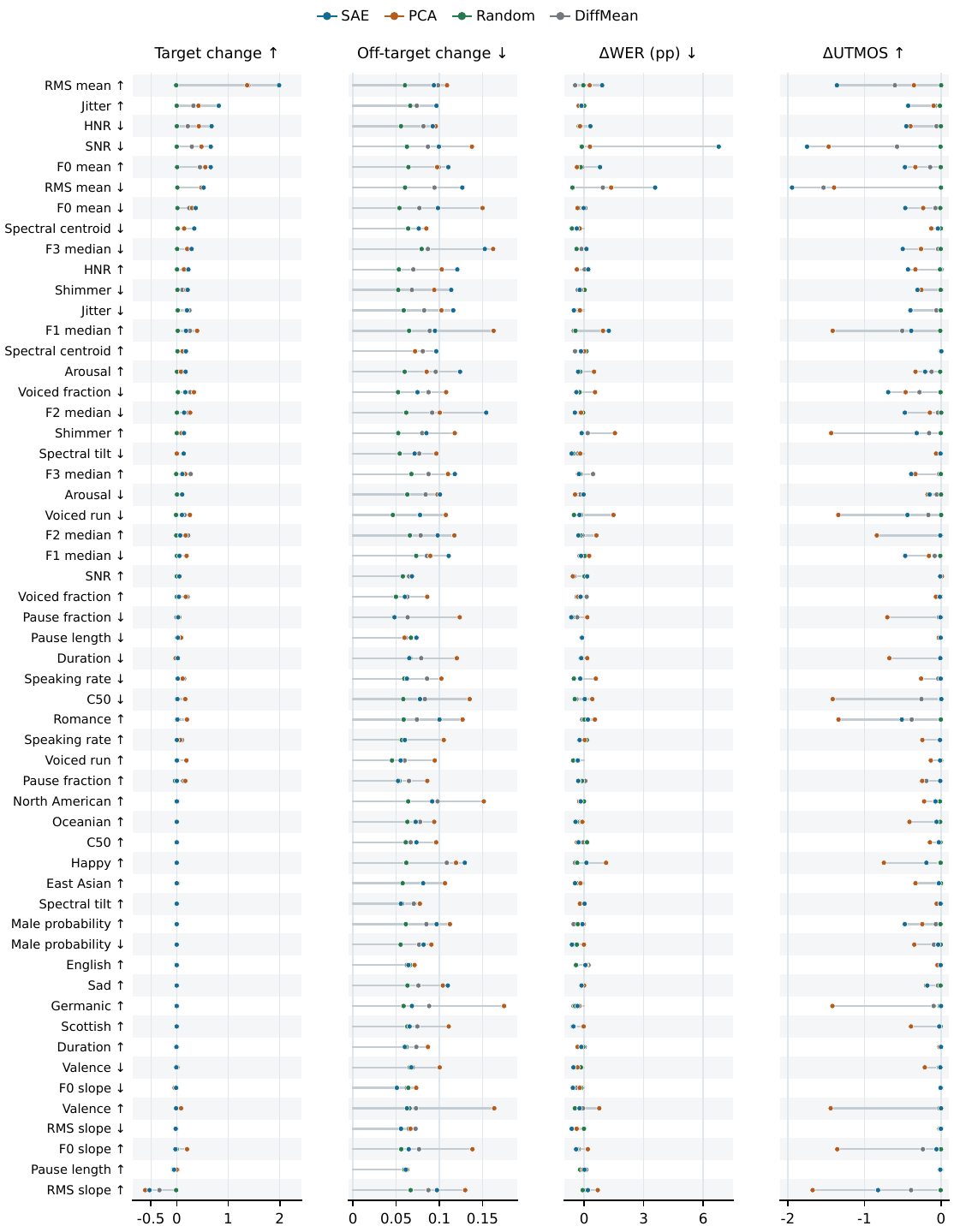}
    \captionof{figure}{Steering results for all $55$ concepts at Qwen3-TTS layer $19$, extending Figure~\ref{fig:steering_concept_lollipop}. Concepts are ordered by descending SAE Target Change.}
    \label{fig:steering_qwen_all55}
\end{minipage}

\noindent\begin{minipage}{\linewidth}
    \centering
    \captionsetup{hypcap=false}
    \includegraphics[width=\linewidth]{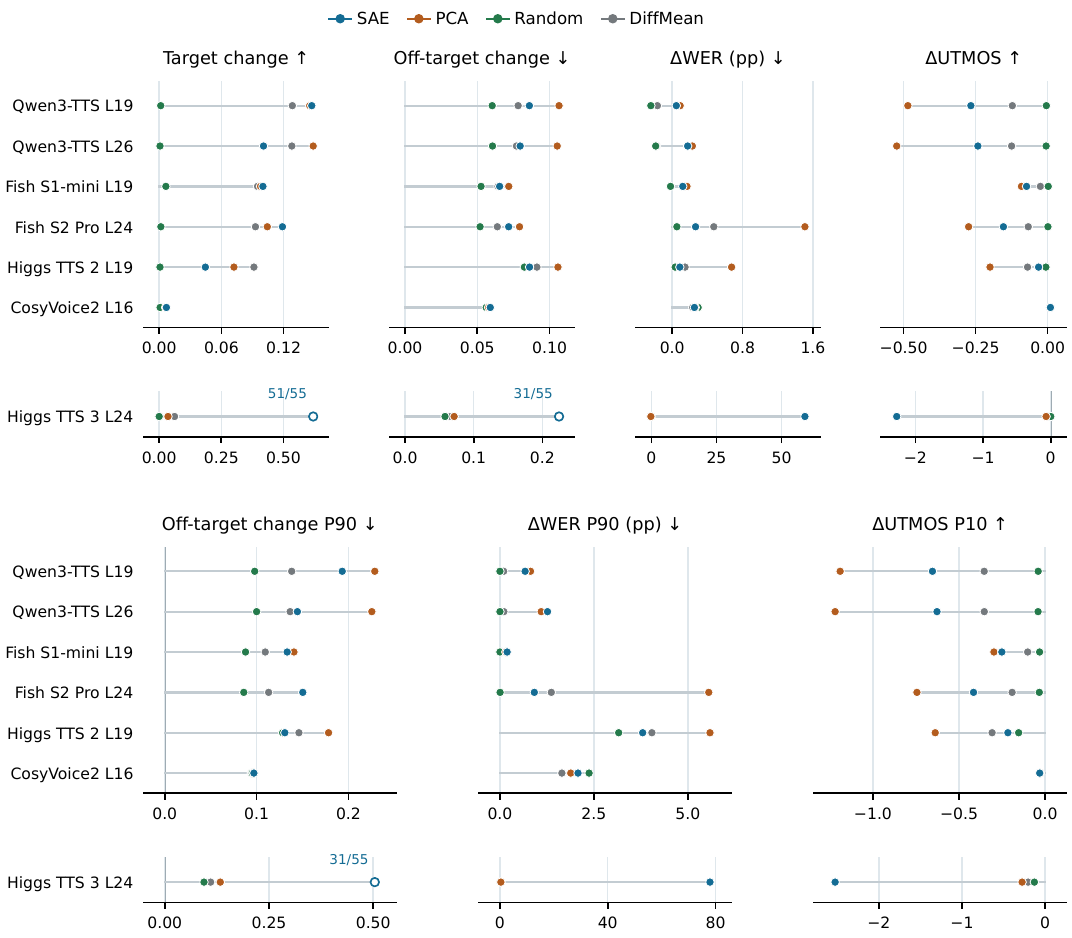}
    \captionof{figure}{Expanded model comparison from Figure~\ref{fig:steering_aggregate}, including Higgs TTS 3 on separate axes. Additional panels show larger off-target changes ($90$th percentile across protected-concept medians) and quality losses ($90$th/$10$th percentiles of utterance-level $\Delta$WER/$\Delta$UTMOS), averaged across target concepts. Hollow markers indicate $51/55$ target and $31/55$ off-target scores for Higgs TTS 3 SAE due to unavailable extractor measurements; all quality means retain $55/55$, including degraded speech.}
    \label{fig:steering_multimodel_extended}
\end{minipage}

\section{Auto-Interpretability Details}
\label{app:autointerp_details}

\para{Feature sampling.}
All experiments use layer-$19$ features on the analysis split, with activating and contrast candidates defined in Section~\ref{sec:novel_concept}. Features must have at least $20$ candidates of each type. We sample $500$ per set from $3583$ eligible SAE features (out of $4096$), $1024$ PCA features, and $4096$ random directions.

\para{Utterance sets.}
Sample counts follow Section~\ref{sec:novel_concept}. Label Fit uses held-out audio; Label Stability repeats the original evidence and uses a disjoint evidence set. Coherence trials randomize clip order. Audio is truncated to at most $20$ seconds.

\para{Validation.}
Features must pass all three thresholds in Section~\ref{sec:novel_concept}. The Coherence threshold of $3$ correct trials out of $5$ is the smallest integer threshold with one-sided binomial probability below $0.05$ under random guessing ($p=\nicefrac{1}{11}$). Stability averages the original-clip and disjoint-clip comparisons. Among the $684$ passing features, these components average $0.849$ and $0.662$, respectively.

\para{Baselines and controls.}
Random-guessing baselines are analytical. The Stability reference assumes uniform guesses over same, overlapping, different, and unclear, giving expected support $0.25\cdot1.0+0.25\cdot0.5=0.375$. Shuffled-label controls assign labels from another feature in the same set and layer, with no feature retaining its own label. Random-audio controls replace each clip with a random utterance. Both retain the original prompts and response formats.

\para{Speaker diversity.}
We cluster speaker embeddings from all strongly activating utterances per feature ($20$--$11778$ for classified features), using average linkage and a cosine-distance cutoff of $0.4$. Of all $4096$ SAE features, $518$ have $1$--$2$ embedding clusters, $181$ have $3$--$5$, $2903$ have more than $5$, and $494$ have fewer than $20$ activating utterances and remain unclassified. The main comparison uses the $3583$ auto-interpretability-eligible features of which $699$ have at most $5$ clusters.

\subsection{Gemini Prompt Templates}
\label{app:prompts}
We use Gemini $3.5$ Flash for all API calls. The exact prompts used for auto-interpretability are below. The request builder interleaves these prompts with text labels and inline audio attachments.

\begingroup
\lstset{emptylines=0, aboveskip=0pt, belowskip=0pt}
\noindent\textbf{Label generation prompt.}\par\nobreak\smallskip
\begin{lstlisting}[basicstyle=\scriptsize\ttfamily, breaklines, columns=fullflexible, keepspaces=true, showstringspaces=false]
I'm studying learned features inside a speech synthesis model. I've attached two
groups of short speech clips for a given feature.

The clips are attached inline in this message. Each audio attachment is
immediately preceded by a text label. Treat these labels as the source of truth:
labels ACTIVATING_1, ACTIVATING_2, ... are Group A, and labels CONTRAST_1,
CONTRAST_2, ... are Group B.

GROUP A (labels ACTIVATING_*): These clips all strongly activate a specific
internal feature of the model.
GROUP B (labels CONTRAST_*): These are random clips that do NOT activate this
feature.

Your task: Listen carefully to both groups and identify what acoustic, vocal, or
recording property consistently distinguishes Group A from Group B.

First, listen for the most salient pattern without assuming it belongs to any
predefined category. Learned model features can correspond to subtle, unusual,
or cross-cutting properties.

After forming your initial hypothesis, use the following list only as a
checklist of possibilities you may have missed, not as a constraint on the
answer:
- Voice quality (breathy, creaky, nasal, hoarse, resonant, thin, etc.)
- Recording conditions (reverb, background noise type, microphone quality, room
  size, phone vs studio)
- Prosody (monotone vs expressive, rising/falling intonation patterns, rhythm)
- Speaking style (read speech vs conversational, formal vs casual, acted vs
  natural)
- Tempo and pacing (fast, slow, many pauses, rushed)
- Pitch register (not just high/low, but placement - head voice, chest voice)
- Energy dynamics (consistently loud, quiet, dynamic range, fading)
- Emotional tone (but be specific - not just "happy" but "politely
  enthusiastic" or "suppressed frustration")
- Articulatory features (mumbling, crisp enunciation, specific vowel qualities,
  accent features)
- Any other pattern you notice, even if subtle, weird, mixed, or hard to name

Be specific and concrete. Don't say "Group A sounds different" - say exactly
how. If you notice multiple distinguishing properties, list them in order of how
consistent they are across all Group A clips. If you're uncertain, say so and
give your best guess with a confidence level.

Prefer the pattern that best separates Group A from Group B, even if it does not
fit cleanly into any category above. It is acceptable for the best explanation
to combine multiple cues, such as "female speaker with close-mic studio audio"
or "fast conversational delivery with clipped endings."

Important: this is a multimodal audio request. Do not answer as if you only have
text labels. If your first instinct is to say you cannot access or play the
audio, double-check the attached inline audio parts and use them. If the provider
truly fails to expose the audio to you, still return the JSON schema below with
confidence "low", explain the access problem in notes_for_human_reviewer, and do
not invent acoustic observations.

Return only one valid JSON object. Do not wrap it in markdown. Use exactly these
keys:
- primary_pattern: one-line description
- confidence: one of "high", "medium", or "low"
- details: 2-3 sentences elaborating the pattern
- secondary_patterns: list of additional distinguishing features, if present
- counter_examples: list noting any Group A clips that don't fit the pattern, or
  Group B clips that do
- distinguishing_cues: list of short concrete acoustic/vocal/recording cues
- label_short: concise lowercase dashboard-friendly label
- is_likely_speaker_identity: boolean
- notes_for_human_reviewer: string
\end{lstlisting}

\noindent\textbf{Coherence prompt.}\par\nobreak\smallskip
\begin{lstlisting}[basicstyle=\scriptsize\ttfamily, breaklines, columns=fullflexible, keepspaces=true, showstringspaces=false]
I'm studying whether a learned speech-model feature captures a coherent audio
concept. I attached several short clips. Most clips belong together because
they strongly activate the same internal feature; exactly one clip is a random
non-activating intruder.

Each audio attachment is immediately preceded by a label CLIP_1, CLIP_2, ...
Listen to all clips and identify the one clip that does NOT belong with the
others. Use the audio content, not the label names.

Return only one valid JSON object with exactly these keys:
- intruder_label: one of the provided CLIP_* labels
- confidence: one of "high", "medium", or "low"
- reason: short explanation of the acoustic/vocal/recording cue that made the
  selected clip seem different
\end{lstlisting}

\noindent\textbf{Label Fit prompt.}\par\nobreak\smallskip
\begin{lstlisting}[basicstyle=\scriptsize\ttfamily, breaklines, columns=fullflexible, keepspaces=true, showstringspaces=false]
I'm evaluating an automatically generated description of a learned speech-model
feature.

Description: "{label}"

Listen to the attached speech clip. Does this clip match the description above?
Answer based on the audio, not on the file name or metadata.

Return only one valid JSON object with exactly these keys:
- answer: "YES" or "NO"
- confidence: one of "high", "medium", or "low"
- reason: short explanation of the acoustic/vocal/recording cue you used
\end{lstlisting}

\noindent\textbf{Label Stability prompt.}\par\nobreak\smallskip
\begin{lstlisting}[basicstyle=\scriptsize\ttfamily, breaklines, columns=fullflexible, keepspaces=true, showstringspaces=false]
I'm evaluating cross-validation consistency for automatic labels of learned
speech-model features.

Each label was generated by an automatic labeling model that compared
activating audio clips for one learned feature against non-activating contrast
clips, then wrote a short primary_pattern description of the acoustic, vocal,
speaking-style, or recording property the feature appears to capture.

The main reported label was generated from the main labeling utterance set and
is the label used for downstream detection scoring. The held-out label was
generated independently from disjoint held-out audio clips for the same feature,
again by comparing activating clips against non-activating contrast clips.

Main reported label:
"{main_label}"

Held-out label ({heldout_name}):
"{heldout_label}"

Decide whether the held-out label supports the same underlying concept as the
main reported label. Treat paraphrases, different wording, and different levels
of specificity as the same if they point to a compatible acoustic/vocal/
recording property. Mark them different if their central claims are
incompatible or mainly describe different properties.

Use this rubric:
- same: The labels refer to substantially the same concept, even if phrased differently.
- overlapping: The labels are compatible but one is broader, narrower, or captures only part of the other.
- different: The labels describe different concepts.
- unclear: There is not enough information to decide.

Return only one valid JSON object with exactly these keys:
- verdict: one of "same", "overlapping", "different", or "unclear"
- support_score: one of 1.0, 0.5, or 0.0
- confidence: one of "high", "medium", or "low"
- reason: one short sentence explaining the judgment
\end{lstlisting}
\endgroup

\FloatBarrier
\section{Additional Results}
\label{app:addlresults}

\subsection{Expanded Concept Metrics}
\label{app:extractor_concept_metrics_appendix}

Figures~\ref{fig:psc_concept_metrics_appendix} and~\ref{fig:extractor_concept_metrics_appendix} show all human-labelled and extractor concepts at Qwen3-TTS layer $19$ for all four methods, ordered by descending SAE all-feature nF1.

\noindent\begin{minipage}{\linewidth}
    \centering
    \captionsetup{hypcap=false}
    \includegraphics[width=0.85\linewidth]{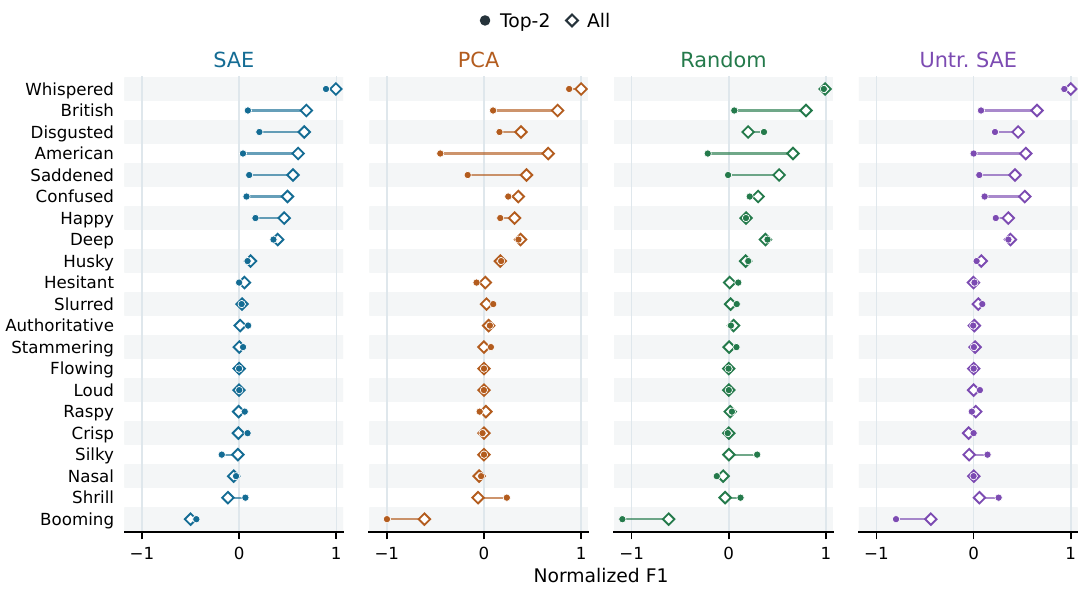}
    \captionof{figure}{All-feature and top-$2$ nF1 for all $21$ human-labelled concepts.}
    \label{fig:psc_concept_metrics_appendix}
\end{minipage}

\newpage
\noindent\begin{minipage}{\linewidth}
    \centering
    \captionsetup{hypcap=false}
    \includegraphics[width=\linewidth]{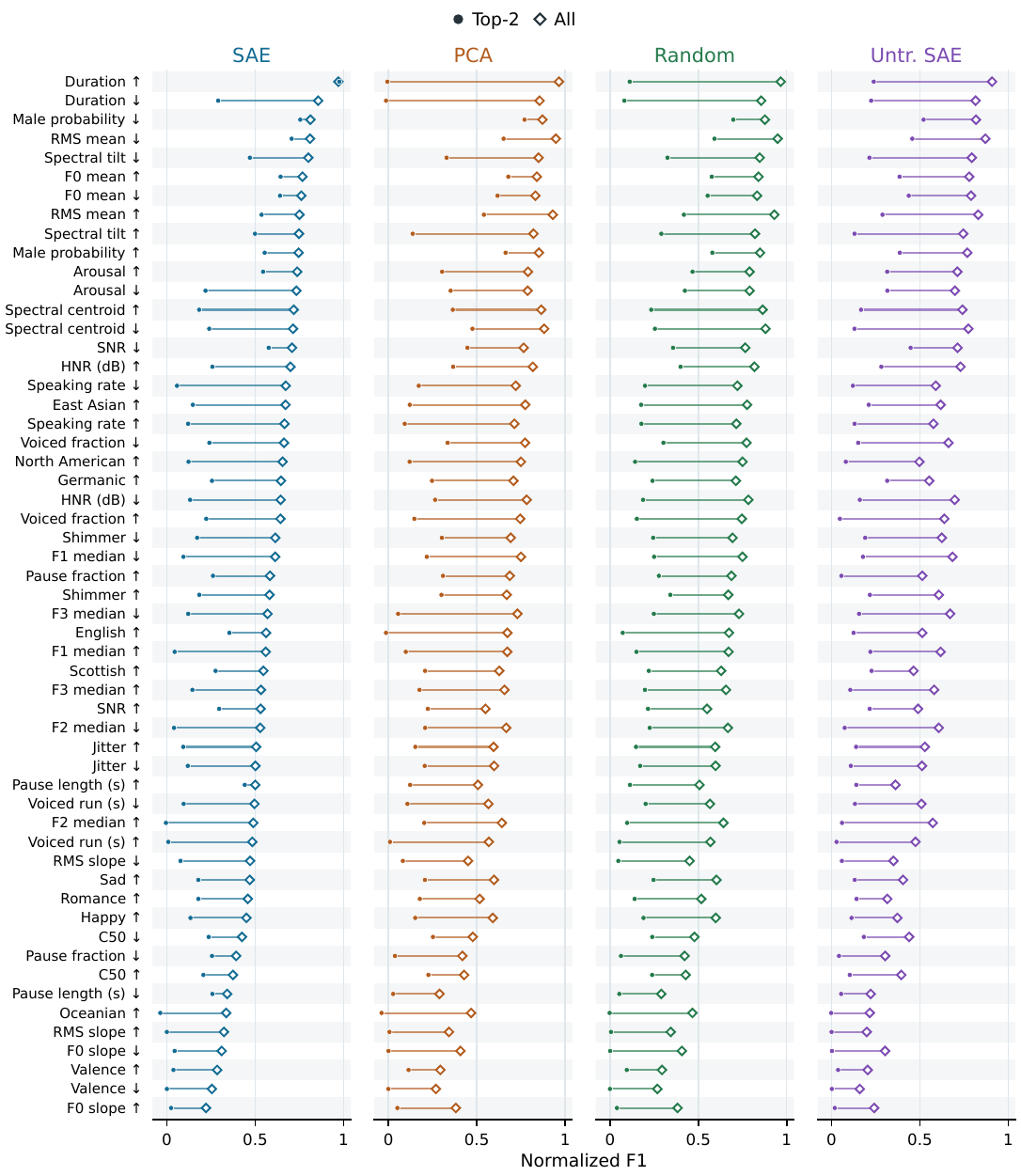}
    \captionof{figure}{All-feature and top-$2$ nF1 for all $55$ extractor concepts at Qwen3-TTS layer $19$, ordered by descending SAE all-feature nF1.}
    \label{fig:extractor_concept_metrics_appendix}
\end{minipage}

\noindent\begin{minipage}{\linewidth}
\subsection{Probing Across Feature Counts}
\label{app:known_concept_k}

    \centering
    \captionsetup{hypcap=false}
    \includegraphics[width=\linewidth]{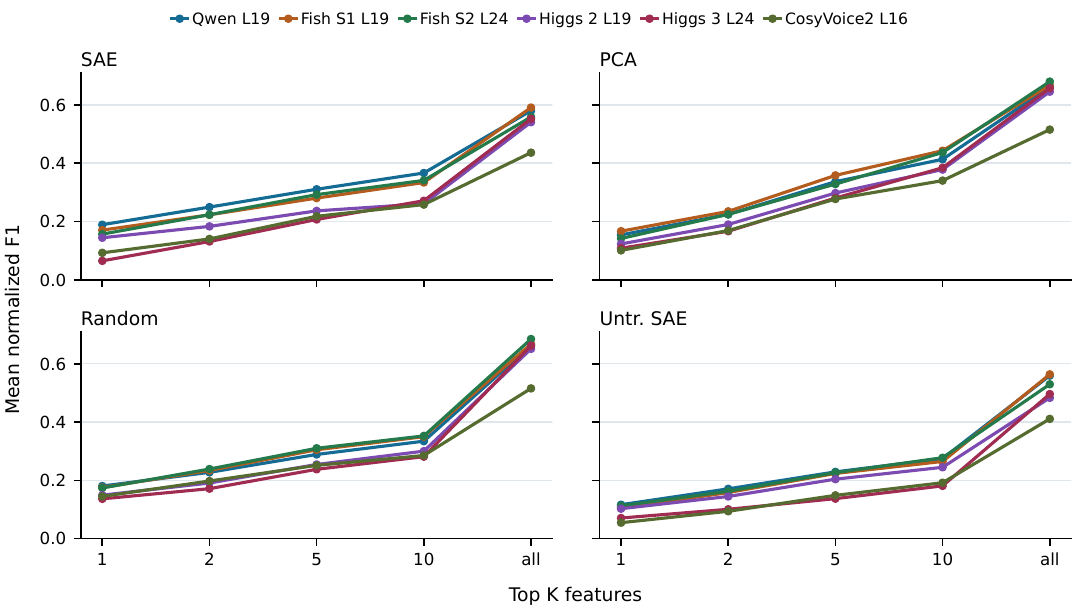}
    \captionof{figure}{Mean nF1 over all $55$ extractor concepts as $k$ increases. \emph{Takeaway}: Extractor prediction improves with more features across all six models and four methods.}
    \label{fig:sparse_probe_by_k}
    \par\medskip
    \includegraphics[width=\linewidth]{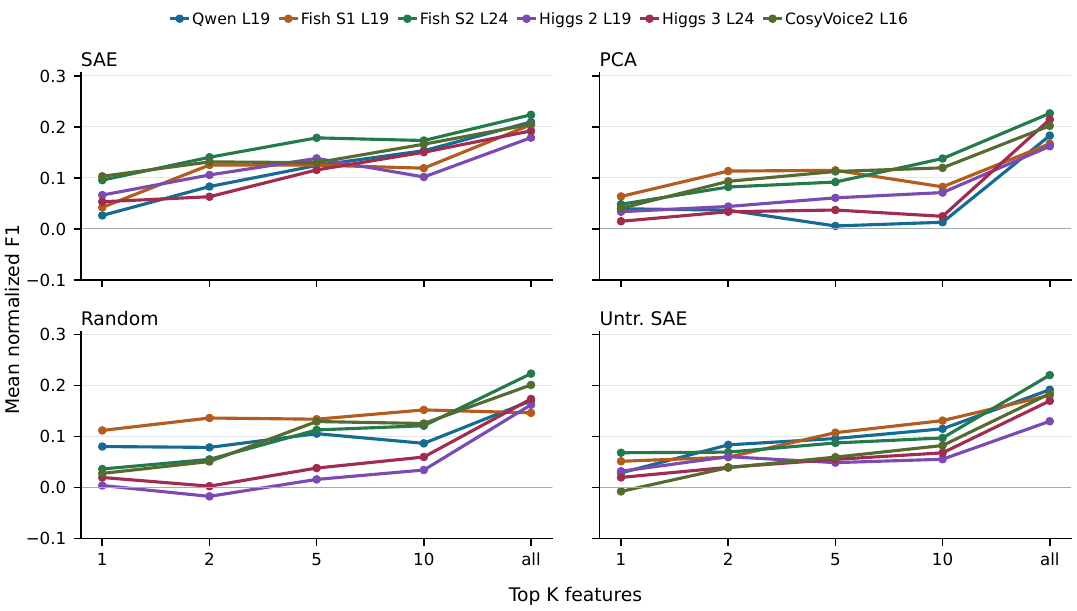}
    \captionof{figure}{Mean nF1 over all $21$ human-labelled concepts. \emph{Takeaway}: Prediction generally improves with more features, with SAE exceeding PCA at $k=2,5,10$ across all six models.}
    \label{fig:psc_sparse_probe_by_k}
\end{minipage}

\noindent\begin{minipage}{\linewidth}
\subsection{Known-Concept Layerwise Trends}
\label{app:known_concept_layerwise}

    \centering
    \captionsetup{hypcap=false}
    {\small Extractor concepts: all features\par}
    \includegraphics[width=0.95\linewidth,trim=-5.123bp 0 0 0,clip]{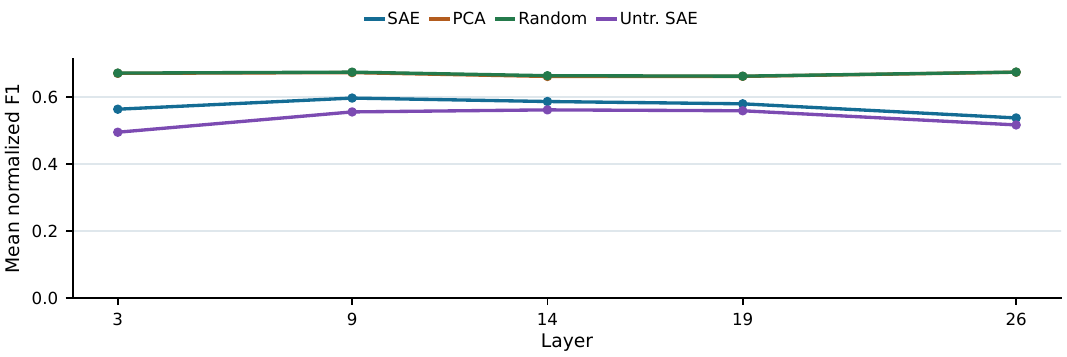}
    \par\smallskip
    {\small Extractor concepts: top-$2$ features\par}
    \includegraphics[width=0.95\linewidth,trim=0 0 0 18bp,clip]{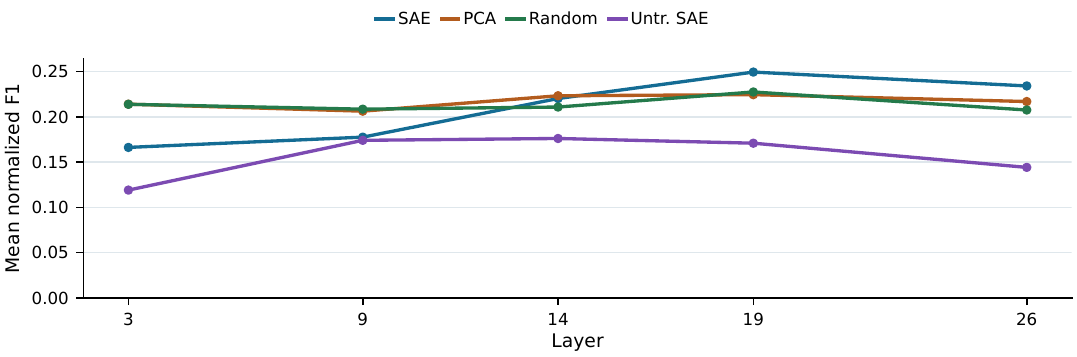}
    \par\smallskip
    {\small Human-labelled concepts: all features\par}
    \includegraphics[width=0.95\linewidth,trim=0 0 0 18bp,clip]{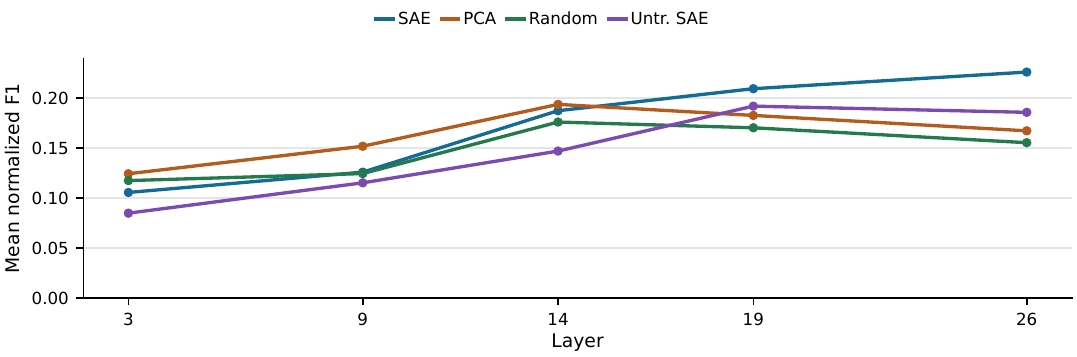}
    \par\smallskip
    {\small Human-labelled concepts: top-$2$ features\par}
    \includegraphics[width=0.95\linewidth,trim=0 0 0 18bp,clip]{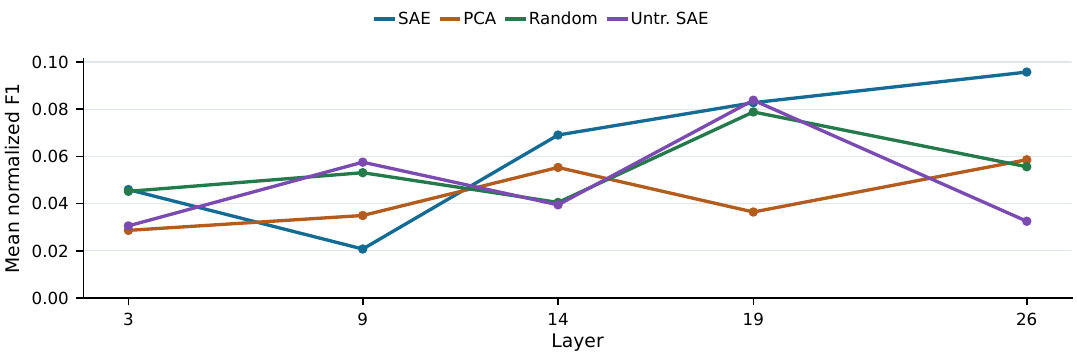}
    \captionof{figure}{Mean nF1 across Qwen3-TTS layers. From top to bottom: extractor all-feature, extractor top-$2$, human-labelled all-feature, and human-labelled top-$2$ scores. \emph{Takeaway}: SAE top-$2$ prediction peaks at layer $19$ for extractor concepts and layer $26$ for human-labelled concepts.}
    \label{fig:known_concept_layerwise}
\end{minipage}

\end{document}